\documentclass[twocolumn]{aastex62}

\received{}
\revised{}
\accepted{}
\usepackage{amssymb}
\usepackage{amsmath}

\def \half {\textstyle{\frac{1}{2}}}

\def \p {\partial}

\def \. {\cdot}

\newcommand{\pder}[2][]{\frac{\partial#1}{\partial#2}}
\newcommand{\kms}{km~s$^{-1}\,$}
\newcommand{\alfven}{Alfv{\'e}n\,}

\begin{document}
\title{A Cancellation Nanoflare Model for Solar Chromospheric and Coronal Heating II. 2D Theory and Simulations}
\correspondingauthor{P. Syntelis}
\email{ps84@st-andrews.ac.uk}

\author{P. Syntelis}
\affiliation{St Andrews University, Mathematics Institute, St Andrews KY16 9SS, UK}

\author{E.R. Priest}
\affiliation{St Andrews University, Mathematics Institute, St Andrews KY16 9SS, UK}

\author{L.P. Chitta}
\affiliation{Max Planck Institute for Solar System Research, G\"{o}ttingen, Germany}

%\shorttitle{Cancellation Nanoflare Model}
%\runningtitle{Cancellation Nanoflare Model}
%\shortauthors{PRIEST, CHITTA}
%\runningauthor{Priest \&  Chitta}

%\institute{$^1$ School of Mathematics and Statistics, University of St. Andrews, Fife KY16 9SS, Scotland, UK\\
%email: \url{erp@st-andrews.ac.uk}\\
%$^2$ Max-Planck-Institut f{\"u}r Sonnensystemforschung, 
%Justus-von-Liebig-Weg 3, D-37077 G{\"o}ttingen, Germany}
%email: \url{dana@solar.physics.montana.edu}\\

\begin{abstract} 

Recent observations at high spatial resolution have shown that magnetic flux cancellation occurs on the solar surface much more frequently than previously thought, and so this led \citet{Priest_etal2018} to propose  magnetic reconnection driven by photospheric  flux cancellation as a mechanism for chromospheric and coronal heating. In particular, they estimated analytically the amount of energy released as heat and the height of the energy release during flux cancellation.
In the present work, we take the next step in the theory by setting up a two-dimensional resistive MHD simulation of two cancelling polarities in the presence of a horizontal external field and a stratified atmosphere in order to check and improve upon the analytical estimates.  Computational evaluation of the energy release during reconnection is found to be in good qualitative agreement with the analytical estimates. In addition, we go further and undertake an initial study of the atmospheric response to  reconnection. We find that, during the cancellation, either hot ejections or cool ones or a combination of both hot and cool ejections can be formed, depending on the height of the reconnection location. The hot structures can have the density and temperature of coronal loops, while the cooler structures are suggestive of surges and large spicules. 

\end{abstract}


\section{Introduction} 
\label{sec:introduction}

The emergence of new magnetic flux from below the photosphere \citep{harvey73} and its reconnection with the overlying magnetic field has long been recognised as being one way of heating part of the solar corona, namely, X-ray bright points \citep{golub74}, and of heating small flares \citep{heyvaerts77}. It has also been proposed as a possible source of coronal X-ray jets \citep{shibata92}, for which there has been a host of observational papers \citep[e.g.,][]{moore10,shimojo00} and numerical experiments  \citep[e.g.,][]{yokoyama96, Archontis_etal2010, morenoinsertis13, Syntelis_etal2015}. Indeed, it  is now appreciated that reconnection can produce a mixture of hot and cold structures and that their origin can be highly subtle and complex \citep[e.g.][]{Hansteen_etal2017,nobrega17,nobrega18}.

The cancellation of photospheric magnetic flux is another common process \citep{martin85} which has been proposed as a mechanism for heating  X-ray bright points  \citep{priest94b,parnell95}, in which magnetic reconnection is driven in the overlying atmosphere during the approach of opposite-polarity magnetic fragments before they actually cancel. Indeed, we shall include this pre-cancellation phase in our use of the words ``flux cancellation". Photospheric flux cancellation has been shown to be associated with both hot and cool jets and also with many different examples of small-scale energy release, such as Ellerman bombs, UV bursts and IRIS bombs \citep[e.g.][]{
Watanabe_etal2011,
Vissers_etal2013,
Peter_etal2014,
Kim_etal2015,Vissers_etal2015,Rutten_etal2015,Rezaei_2015,
Nelson_etal2016,Tian_etal2016,Reid_etal2016,Rutten_2016,
Nelson_etal2017,Toriumi_etal2017,Hong_etal2017,Libbrecht_etal2017,RouppevanderVoort2017}. 

A new achievement is the remarkable observations from the Sunrise balloon mission \citep{solanki10a,solanki17b}, which have revealed images of the photospheric magnetic field at a spatial resolution of 0.15 arcsec, which is six times better than the Helioseismic Imager (HMI) on the Solar Dynamics Observatory (SDO). In particular, they show that magnetic flux is emerging and cancelling at a rate of 1100 Mx cm$^{-2}$ day$^{-1}$ \citep{smitha17}, which is an order of magnitude higher than previously realised. Furthermore, whereas, at the spatial resolution of HMI, coronal loops have their footpoints located in regions of uniform polarity, at Sunrise resolution  the footpoints are surprisingly revealed to have mixed polarity that is cancelling at a rate of 10$^{15}$ Mx s$^{-1}$ \citep{chitta17b}. Other examples of flux cancellation producing coronal loop brightening have been presented by \citet{tiwari14}, \citet{huang18}, and \citet{chitta18}

These observations led \citet{Priest_etal2018} to propose reconnection driven by photospheric flux cancellation as a mechanism for heating the chromosphere and corona.  They set up an analytical model for the approach and cancellation of two opposite-polarity magnetic fragments of flux $\pm F$ in the photosphere in the presence of an overlying uniform horizontal magnetic field $B_0$, and found that the evolution of the system depends on the value of  a key parameter, called the {\it interaction\ distance}, which, for three-dimensional sources, may be written as
\begin{equation}
d_0^{3D}=\left( \frac{F}{\pi B_0}\right)^{1/2}.
\end{equation}

Suppose the magnetic flux sources are separated by a distance $2d$. Then, when $d>d_0^{3D}$, the sources are not connected magnetically, but when $d=d_0^{3D}$ a null point (or in 3D a {\it separator}) forms in the photosphere. As the sources approach closer, such that $d<d_0^{3D}$, reconnection is driven and the reconnection location rises in the atmosphere to a maximum height proportional to $d_0^{3D}$.  Thereafter, the reconnection location moves back towards the solar surface, which it reaches when the two sources come into contact and cancel ($d=0$).  Thus, the maximum reconnection height can be located in the photosphere or chromosphere if $d_0^{3D}$ is small enough or in the corona if it is large enough.  

As well as calculate the way the reconnection height depends on flux ($F$) and overlying field strength ($B_0$) through $d_0^{3D}$, \citet{Priest_etal2018} made estimates for the energy release, and found that, for reasonable values of the parameters, the heating rate is sufficient to heat the chromosphere and corona.

In the present paper, we develop the model further by setting up a two-dimensional computational experiment for flux cancellation that has the same features as our analytical model, namely, two approaching flux sources in the presence of an overlying horizontal magnetic field, so that we can test the predictions of the analytical model. However, we add an extra feature, namely, a simple stratified atmosphere in order to understand some of the effects of stratification.  

Section \ref{sec:theory} presents some more details of the theory of reconnection in two dimensions, including Sweet-Parker reconnection, fast reconnection, and energy conversion. Then, Section \ref{sec:numerical_simulations} presents our computational model and compares it with the analytical theory, before a summary discussion is given in the final Section.

\section{Theory for Energy Release driven by Photospheric Flux Cancellation in 2D} 
\label{sec:theory}
Here we make some theoretical estimates of the energy release by 
steady-state magnetic reconnection in two dimensions, developing 
the basic theory from \citet{Priest_2014}   in new ways. We will start by briefly describing slow Sweet-Parker reconnection and fast reconnection, and then discussing reconnection driven by magnetic flux cancellation.

\subsection{Slow Sweet-Parker Reconnection}
\label{sec:cartoon_sp_fast}

\begin{figure}
\begin{center}
\includegraphics[width=\columnwidth]{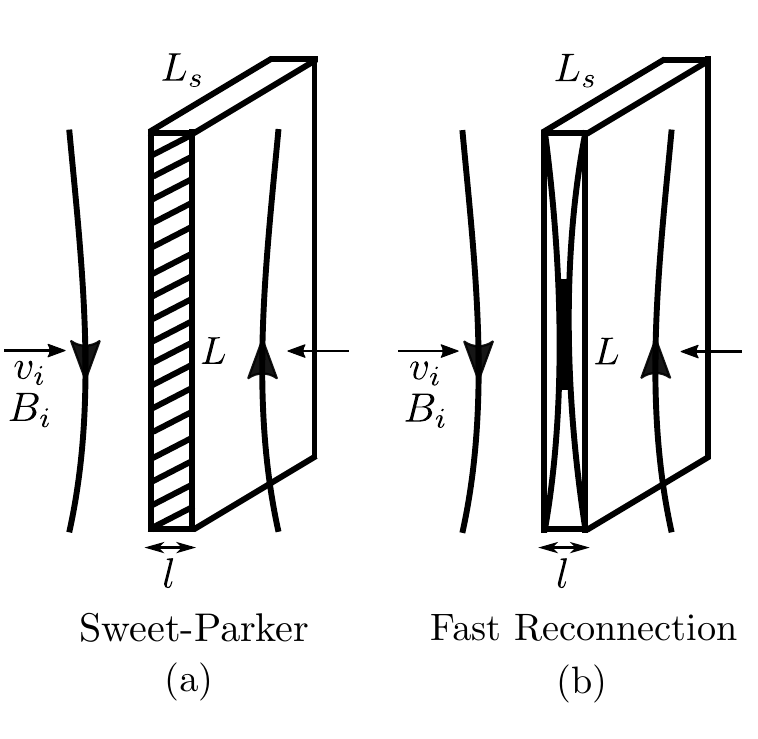}
\caption{
The nomenclature for energy release in a simple 
reconnection region consisting of either (a) a slow Sweet-Parker 
current sheet or (b) a fast reconnection region with a small sheet and four slow-mode shock waves.}
\label{fig:cartoon_sp_fast}
\end{center}
\end{figure}

Consider first a simple Sweet-Parker current sheet of given length $L$, depth $L_{s}$ and width $l$ situated between 
oppositely directed magnetic fields $B_i$ and $-B_i$ (Fig. \ref{fig:cartoon_sp_fast}a). If plasma and magnetic field is brought in from both 
sides at a speed $v_{i}$, then a balance between inwards advection and outwards diffusion of
magnetic flux implies
\begin{equation}
    v_{i}=\frac{\eta}{l}.
\label{eqn1}
\end{equation}
Furthermore, if the plasma has uniform density $\rho_i$, balancing 
the rates of inflow and outflow of mass gives 
\begin{equation}
    L v_{i}=l v_{Ai},
\label{eqn2}
\end{equation}
where $v_{Ai}=B_{i}/\sqrt{\mu \rho_i}$ is the outflow speed from the current sheet, namely, 
the \alfven speed based on the inflow magnetic field.

Eliminating $l$ between Equation (\ref{eqn1}) and (\ref{eqn2}) produces an expression for the
dimensionless inflow speed or \alfven Mach number
($M_{Ai}=v_{i}/v_{Ai}$), i.e., the {\it reconnection rate}, of
\begin{equation}
    M_{Ai}=\frac{1}{R_{mi}^{{1/2}}},
\label{eqn3}
\end{equation}
where $R_{mi}=Lv_{Ai}/\eta$ is the external magnetic Reynolds 
number based on the global external length-scale ($L$) and \alfven 
speed ($v_{Ai}$).  For a given current-sheet length  ($L$) and 
external magnetic field ($B_{i}$),  Equation (\ref{eqn3}) thus provides 
the  Sweet-Parker rate.

Half of the magnetic energy that comes into the reconnection 
region from both sides is converted into heat and the other half into kinetic energy (which can later itself dissipate viscously or through shock waves). The rate of inflow of magnetic energy from one side through an area of  $L L_s$ is just the Poynting influx 
$(E H_{i}L L_s = E B_{i}L L_s/\mu)$, where the magnitude of 
the electric field is $E=v_{i}B_{i}$, and so the rate of conversion to heat of magnetic energy coming in from both sides of the current sheet is
\begin{equation}
    \frac{dW}{dt} = \frac{v_{i}B_{i}^{2}}{\mu}L L_s= 
    \frac{1}{R_{me}^{{1/2}}}\frac{v_{Ai}B_{i}^{2}}{\mu}L L_s,
\label{eqn4}
\end{equation}
after substituting for $v_{i}$ from Equation (\ref{eqn3}). The phrase ``two-dimensional" can refer to a situation in which the variables are situated in three dimensions but they depend on only two of them, such as $x$ and $y$, but it can also refer to variables that exist only in two dimensions, in which case the above expression would need to be divided by $L_s$. In what follows it should be clear which of the two definitions is being inferred.

\subsection{Fast Reconnection}

Next, suppose the inflow speed is faster than $v_{Ai}/R_{me}^{1/2}$, 
while the inflow magnetic field ($B_{i}$) and area $L_s$ are 
the same as before (Fig.~\ref{fig:cartoon_sp_fast}b). Then, three possibilities have been studied.
Firstly, according to {\it fast steady-state reconnection} theory 
(either \citet{petschek64} or Almost-Uniform \citep{priest86a}), the reconnection region possesses a 
complex internal structure consisting of a central small Sweet-Parker current 
sheet together with four slow-mode shock waves propagating from their 
ends and standing in the flow. As the speed increases, the central 
 sheet diminishes in size, while the length and 
inclination of the shock waves increases. Most of the energy 
conversion then takes place at the shock waves, with $\frac{2}{5}$ of the 
inflowing magnetic energy being converted to heat (rather than the $\half$ that is found in Sweet-Parker reconnection) and the 
remainder going to kinetic energy.

Secondly, {\it fast collisionless reconnection} is helped by the Hall effect, 
when the resistive diffusion region is replaced by an ion diffusion 
region and a smaller electron diffusion region. In this case, a similar
fast maximum rate of reconnection as in Petschek's mechanism results 
\citep{shay98a,birn01,huba03,huba04,shay07,birn07}.

Thirdly, when the central sheet is long enough, it goes unstable to 
secondary tearing mode instability and a regime of {\it impulsive bursty 
reconnection} results, first described by 
\citet{priest86b,lee86a,biskamp86,forbes87} and later studied by 
\citet{loureiro07,loureiro12,loureiro13,bhattacharjee09}.  Reconnection is then fast but time-dependent and impulsive, although the mean rate is likely to be 
similar to the previous cases.
 
For each of the three above scenarios, Equation (\ref{eqn1}) no 
longer holds, but the same mass conservation relation holds as before for the reconnection region as a 
whole, namely, 
\begin{equation}
    Lv_{i}=l v_{Ai},  
\label{eqn5}
\end{equation}
where $L$ now refers to the length of the whole reconnection region (including shock waves and central current sheet) rather than the length of just the central sheet, and variables with subscript $i$ refer to values at the inflow to that whole region.
Equation (\ref{eqn5}) determines the overall width ($l$) of the complex 
reconnection region for 
a given $L$, $v_{i}$ and $v_{Ai}$.  The conversion rate of inflowing energy from both sides of the current sheet then becomes
\begin{equation}
    \frac{dW}{dt} = \frac{4}{5}\frac{v_{i}B_{i}^{2}}{\mu}L L_s,  
\label{eqn6}
\end{equation}
where $v_{i}$ possesses any value up to a maximum of typically 
0.01--0.1 of the Alfv\'en speed ($v_{Ai}$).

\subsection{Energy Conversion during Photospheric Flux Cancellation in 2D} 
\label{sect_4}

\subsubsection{Magnetic configuration}
Consider sources of  positive and negative  photospheric magnetic flux ($\pm F$) situated at points  B $(d,0)$ and A $(-d,0)$ on the $x$-axis  in a region of 
uniform magnetic field $B_{0}\bf{\hat{x}}$, and suppose they approach one another at 
speeds $\pm v_{0}$. 

The resulting magnetic field (in two dimensions) above the 
photosphere ($y>0$) is given by
\begin{equation}
{\bf B}=\frac{F}{\pi}\frac{{\bf \hat{r}_{1}}}{r_{1}}-\frac{F}{\pi}\frac{\bf{\hat{r}_{2}}}{r_{2}}+B_{0}{\bf{\hat{x}}},
\label{eqn23}
\end{equation}
where
\begin{equation}
{\bf r}_{1}=(x-d){\bf \hat{x}}+y{\bf \hat{y}}, \ \ \ \ \ \ \ 
{\bf r}_{2}=(x+d){\bf \hat{x}}+y{\bf\hat{y}}, \nonumber
\label{eqn24}
\end{equation}
are the vector distances from the two sources to a point P($x,y$).

\begin{figure*}
\begin{center}
\includegraphics[width=0.7\linewidth]{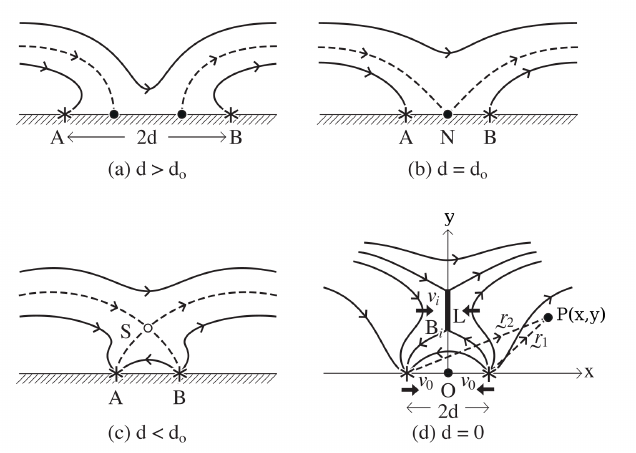}
\caption{Magnetic topology during reconnection driven by photospheric flux cancellation when (a) $d>d_0$, (b) $d=d_0$ and (c) $d<d_0$, where $d$ is half the separation distance of the two flux sources and $d_0$ is the flux interaction distance. (d) The notation used for the reconnection region.}
\label{fig:cartoon}
\end{center}
\end{figure*}

It is natural in Equation~(\ref{eqn23}) to non-dimensionalise the magnetic
field with respect to $B_{0}$ and distances with respect to the 2D version of the interaction distance \citep{Longcope_1998}, namely,
\begin{equation}
d_{0}=\frac{2F}{\pi B_{0}},  
\label{eqn25}
\end{equation}
and so define
\begin{equation}
{\bar B_{x}}=\frac{B_{x}}{B_{0}},\ \ \ \ \ 
{\bar{d}}=\frac{d}{d_{0}},\ \ \ \ \ {\bar{y}}=\frac{y}{d_{0}}. 
\nonumber
\label{eqn26}
\end{equation}
Then the magnetic field on the $y$-axis becomes
\begin{equation}
{\bar B_{x}}=\left(-\frac{{\bar d}}{{\bar y}^{2}+{\bar d}^{2}}+1\right).
\label{eqn27}
\end{equation}

\begin{figure}
\begin{center}
\includegraphics[width=\columnwidth]{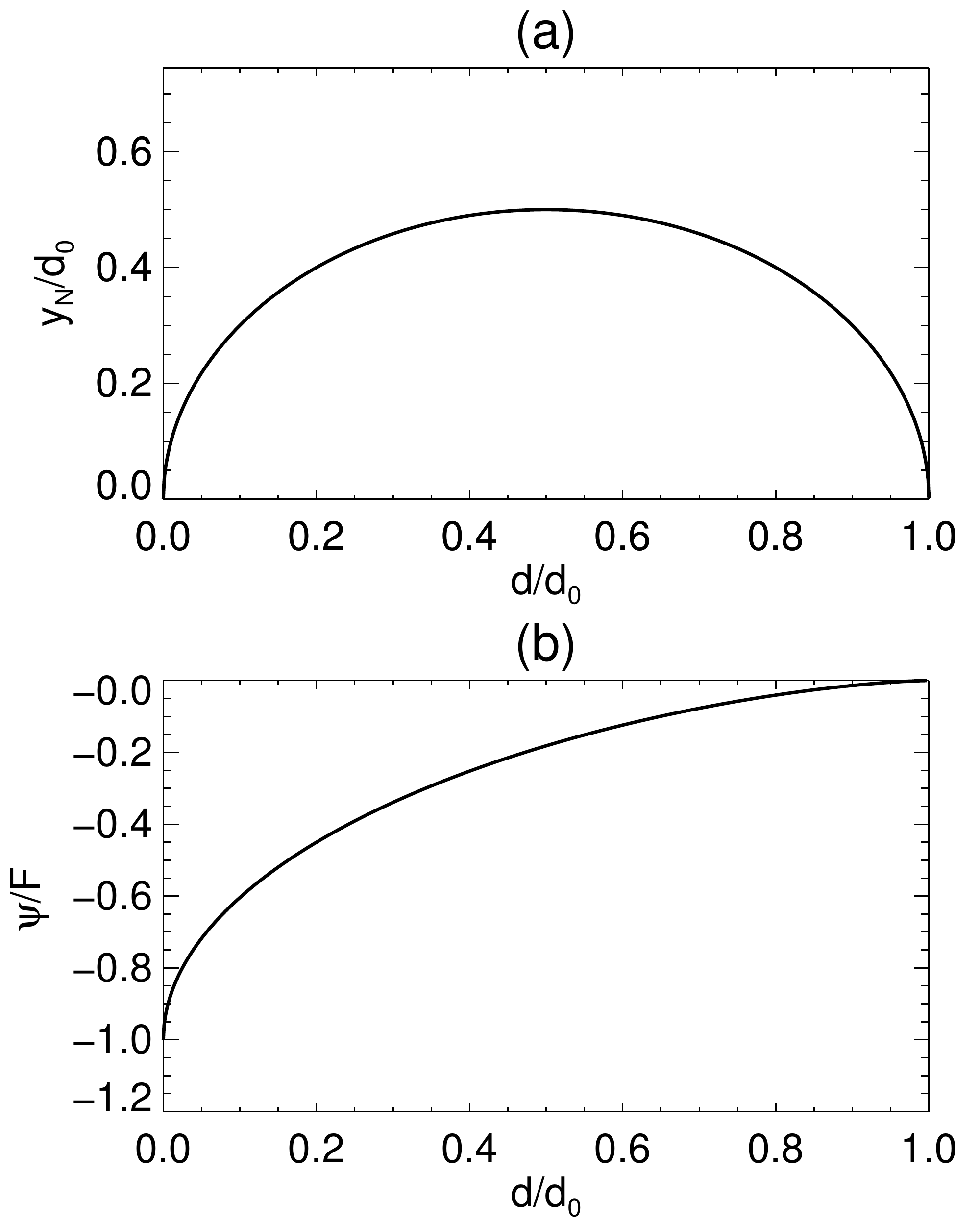}
\caption{
(a) The height of the null point ($y_N$) given by Equation~(\ref{eqn28}) as a function of the distance ($d$) of the sources from the origin (as shown in Fig.~\ref{fig:cartoon}c). 
(b) The magnetic flux ($\psi$) below the null given by Equation~(\ref{eqn33}).}
\label{fig:null_flux}
\end{center}
\end{figure}

Consider what happens as the two sources approach each other.
The evolution of the topology  is similar to what happens in three dimensions, as described in detail in Sec. 2.1 of  \citet{Priest_etal2018}. When the two sources are far away ($d>d_0$), they are not connected magnetically and two first-order null points lie on the $x$-axis between the sources (Fig.~\ref{fig:cartoon}a). When $d=d_{0}$, a high-order null point appears at the origin (Fig.~\ref{fig:cartoon}b). As the sources approach one another ($d<d_0$), the null point rises above the photosphere (Fig.~\ref{fig:cartoon}c).
The location of the null point at $y=y_{N}$ is given by 
\begin{equation}
{\bar y_{N}}=\sqrt{{\bar d}-{\bar d}^{2}},
\nonumber
\label{eqn28}
\end{equation}
and it rises along the $y$-axis to a maximum of $y=\half d_{0}$ when 
$d=\half d_{0}$. Then, it falls, reaching the origin when $d=0$, 
as shown in Fig.~\ref{fig:null_flux}a. When $d=0$, the flux of the two sources has completely cancelled.

\subsubsection{Inflowing Plasma Speed ($v_i$) and Magnetic Field ($B_i$) at the Reconnection Site}

To analyze the energy release during flux cancellation, the natural parameters, for each value of the 
source separation ($2d$), are the magnetic diffusivity ($\eta$), the critical source 
half-separation distance ($d_{0}$), the flux source speed ($v_{0} = 
{\dot d}=dd/dt$) and the overlying field strength ($B_{0}$). We now proceed to calculate the inflow 
speed ($v_{i}$) and magnetic field ($B_{i}$) to the current sheet and
the sheet length ($L$) as functions of these parameters in the cases 
of slow reconnection and fast reconnection. The magnetic configuration driven by  flux cancellation is shown in Fig.~\ref{fig:cartoon}d.

First we consider $B_i$. The components of the potential magnetic field sufficiently close to a 2D X-point can be written as \citep{Priest_2014}
\begin{equation}
    B_{y}+iB_{x}=kz, \nonumber
\label{eqn7}
\end{equation}
where $k$ is a constant and $z=x+iy$ is the complex variable.  Suppose that the 
configuration with a reconnecting current sheet of length $L$ is 
represented by 
\begin{equation}
    B_{y}+iB_{x}=k(z^{2}+{\textstyle{\frac{1}{4}}}L^{2})^{1/2},
\label{eqn8}
\end{equation}
such that the sheet is a cut in the complex plane between $z=\pm {\textstyle{\frac{1}{2}}}iL$.
Then, putting $z=0+$ implies that
\begin{equation}
    B_{i}={\textstyle{\frac{1}{2}}}kL, \nonumber
\label{eqn9}
\end{equation}
which is the required expression for $B_i$
when the $x$-component of the field in the potential state near the null has the form $B_x=ky$.
The value of $k$ is calculated as follows.
The horizontal field $B_{x}$ near $y=y_{N}$ may be obtained by putting $y=y_{N}(1+\epsilon)$, where 
$\epsilon\ll 1$ in Equation (\ref{eqn27}). Keeping only the 
linear terms, this equation gives
\begin{equation}
{\bar B_{x}}=2\epsilon(1-{\bar d}),
\nonumber
\label{eqn29}
\end{equation}
or
\begin{equation}
B_{x}=2\sqrt{\frac{{d_{0}-d}}{d}}\left(\frac{y-y_{N}}{d_{0}}\right)B_{0}.
\label{eqn30}
\end{equation}
This determines the value of $k$, and so our required expression becomes
\begin{equation}
\frac{B_{i}}{B_{0}}=\sqrt{\frac{d_{0}}{d}-1}\frac{L}{d_{0}}.   
\label{eq:bi}
\end{equation}

Next, consider $v_{i}$. This may be calculated from the rate of 
change (${\dot \psi} \equiv d\psi /dt$) of magnetic flux, since
\begin{equation}
  v_{i}B_{i}={\dot \psi},
  \label{eq:vibi}
\end{equation}
or, in dimensionless form
\begin{equation}
  \frac{v_{i}}{v_{A0}}=\frac{{\dot 
  \psi}}{v_{A0}B_{0}}\frac{B_{0}}{B_{i}}, 
  \label{eqn32}
\end{equation}
where $v_{A0}$ a hybrid Alfv\'en speed based on the magnetic field 
$B_{0}$ and the density of the inflowing material, namely,
\begin{equation}
v_{A0} = \frac{B_0}{\sqrt{\mu \rho_i}}.
\label{eq:hybrid_va}
\end{equation}
In turn, ${\dot \psi}$ may be calculated from the reconnected flux 
($\psi$), as estimated from the magnetic flux below the 
null point, namely,
\begin{align}
\psi &=\int_{0}^{y_{N}}B_{0}-\frac{2Fd}{\pi(y^{2}+d^{2})}dy \nonumber \\
&=\frac{2F}{\pi} \int_{0}^{\bar{y_{N}}}1-\frac{\bar{d}}{({\bar 
y}^{2}+{\bar{d}}^{2})}d{\bar y} \nonumber \\
&=\frac{2F}{\pi}\left(\sqrt{\bar d - {\bar 
d}^{2}}-\tan^{-1}\frac{\sqrt{1-\bar d}}{\sqrt{\bar d}}\right).
\label{eqn33}
\end{align}
It can be seen from Fig.~\ref{fig:null_flux}b that, as expected, the reconnected flux 
vanishes when $d=d_{0}$ and increases monotonically to a value of $F$ 
as the separation ($2d$) between the sources approaches zero. 

Then,  differentiating Equation (\ref{eqn33}) with respect to time 
determines ${\dot \psi}$ in terms of ${\dot d} = v_{0}$, and Equation (\ref{eqn32}) becomes
\begin{equation}
  v_{i}=v_{0}\frac{d_{0}}{L}. 
  \label{eq:vi}
\end{equation}

\subsubsection{Energy Release}

The rate of inflow of magnetic energy from one side of a current sheet of length $L$, at speed $v_i$, with field strength $B_i$ and density $\rho_i$ is the Poynting flux through that surface. In 2D the surface of the current sheet will be a line of length $L$, so the Poynting influx is $E H_i L = E B_i L / \mu$. Since the electric field is $E=v_i B_i$, and the magnetic energy inflow occurs from both sides of the current sheet, the Poynting influx from both sides will be
\begin{equation}
    S_i = 2 \frac{v_{i}B_{i}^{2}}{\mu}L
\label{eq:poynting}    
\end{equation}
This has units of energy/time/length, since we assume here a purely 2D configuration with no depth in the third dimension.
To derive the energy release, the length of the current sheet and the conversion rate to heat has to be estimated. Both will depend on the type of reconnection (Sweet-Parker or fast). For a configuration with depth $L_S$ in the third dimension this would be multiplied by $L_S$.
\subsubsection{Slow Sweet-Parker Reconnection}
Here we  calculate the energy release for Sweet-Parker reconnection.
After eliminating  $l$ between the Sweet-Parker relations (Equations (\ref{eqn1}) and (\ref{eqn2})) 
we find that the current sheet length is
\begin{equation}
    L=\frac{\eta v_{Ai}}{v_{i}^{2}},  
\label{eqn15}
\end{equation}
which can be  nondimensionalised in terms of $d_{0}$ to give
\begin{equation}
 \frac{L}{d_{0}}=\frac{1}{R_{m0}}\frac{B_{i}}{B_{0}}\frac{v_{A0}^{2}}{v_{i}^{2}},
  \nonumber
  \label{eqn36}
\end{equation}
where $R_{m0}=d_{0}v_{A0}/\eta$ is the magnetic Reynolds number based 
on $d_{0}$ and $v_{A0}$.  We then substitute for $B_{i}/B_{0}$ from 
Equation (\ref{eq:bi}) and $v_{i}$ from Equation (\ref{eq:vi}) to give
\begin{equation}
 \frac{L_{SP}^{2}}{d_{0}^{2}}=R_{m0}\frac{v_{0}^{2}}{v_{A0}^{2}}\frac{1}{\sqrt{d_{0}/d-1}}.
  \label{eq:lsp}
\end{equation}
where the subscript $SP$ denotes the Sweet-Parker current sheet length. Finally, by substituting in Equation (\ref{eq:poynting}) the values of  $v_{i}$ (Equation (\ref{eq:vi})), $B_{i}$ (Equation (\ref{eq:bi})) and $L=L_{SP}$ (Equation (\ref{eq:lsp})), the rate of 
 Poynting influx becomes
\begin{equation}
    S_{i_{SP}}=2\frac{v_{0} B_{0}^{2}}{\mu}d_{0}\sqrt{d_{0}/d-1}\ R_{m0}M_{A0}^{2},
\end{equation}
in terms of the \alfven Mach number 
($M_{A0}=v_{0}/v_{A0}$) based on the flux source speed $v_{0}$.
Since half of the magnetic energy is converted to heat during Sweet-Parker reconnection, the energy release rate will be:
\begin{equation}
    \frac{dW_{SP}}{dt}=\frac{v_{0} B_{0}^{2}}{\mu}d_{0}\sqrt{d_{0}/d-1}\ R_{m0}M_{A0}^{2},
\end{equation}
which has units of energy/time/length for our 2D theory.

\subsection{Fast Reconnection}

We  derive now the energy release for fast reconnection driven by flux cancellation.
During fast reconnection, the length of the current sheet is much smaller than the Sweet-Parker one. $L$ is determined by assuming the inflow speed $v_{i}=\alpha v_{Ai}$. By writing $v_{i}=\alpha v_{Ai} = \alpha v_{A0}B_i/B_0$ and
then using Equation (\ref{eq:bi}) and (\ref{eq:vi}), $L$ becomes
\begin{equation}
    \frac{L^{2}}{d_{0}^{2}}=\frac{v_{0}}{\alpha 
    v_{A0}}\frac{1}{\sqrt{d_{0}/d-1}}.
\label{eq:lfast}
\end{equation}
Then, after substituting for $v_{i}$, $B_{i}$ and $L$ into Equation (\ref{eq:poynting}), we find the rate of energy inflow for fast reconnection as
\begin{equation}
    S_i=2\frac{v_{0} B_{0}^{2}}{\mu}d_{0}\ 
    \sqrt{d_{0}/d-1}\frac{M_{A0}}{\alpha}.
\label{eq:poyntingfast}
\end{equation}
During fast reconnection, $\frac{2}{5}$ of the magnetic energy is converted to heat and  $\frac{3}{5}$ to kinetic energy. Therefore, the rates of kinetic energy release and energy release as heat become
\begin{equation}
    \frac{dK}{dt}=1.2\frac{v_{0} B_{0}^{2}}{\mu}d_{0}\ 
    \sqrt{d_{0}/d-1}\frac{M_{A0}}{\alpha}
\label{eq:kinetic}
\end{equation}
and 
\begin{equation}
    \frac{dW}{dt}=0.8\frac{v_{0} B_{0}^{2}}{\mu}d_{0}\ 
    \sqrt{d_{0}/d-1}\frac{M_{A0}}{\alpha}.
\label{eq:heat}
\end{equation}

\section{Numerical Computations} 
\label{sec:numerical_simulations}

\subsection{Numerical Setup}
\label{sec:model}

To perform the computations, we numerically solve the 2D MHD equations in Cartesian geometry using the Lare3D code (v3.2) of \citet{Arber_etal2001}. The equations in dimensionless form are:
\begin{align}
&\frac{\partial \rho}{\partial t}+ \nabla \cdot (\rho \mathbf{v})  =0 ,\\
&\frac{\partial (\rho \mathbf{v})}{\partial t}  = - \nabla \cdot (\rho \mathbf{v v})  + (\nabla \times \mathbf{B}) \times \mathbf{B} - \nabla P + \rho \mathbf{g} , \\
&\frac{ \partial ( \rho \epsilon )}{\partial t} = - \nabla \cdot (\rho \epsilon \mathbf{v}) -P \nabla \cdot \mathbf{v}+ Q_\mathrm{j}+ Q_\mathrm{v} + Q_c, \\
&\frac{\partial \mathbf{B}}{\partial t} =\nabla \times (\mathbf{v}\times \mathbf{B})- \nabla \times( \eta \nabla \times \mathbf{B}),\\
&\epsilon  =\frac{P}{(\gamma -1)\rho}, \\
&P = \frac{\rho k_B T}{\mu_m},
\end{align}
where $\rho$, $\mathbf{v}$, $\mathbf{B}$ and P are density, velocity vector, magnetic field vector and gas pressure. Gravity is $g_0=274$~m s$^{-1}$. We assume a perfect gas with specific heat of $\gamma=5/3$. Viscous heating ($Q_\mathrm{v}$) and Joule dissipation ($Q_\mathrm{j}$) are included. Heat conduction ($Q_c$) is treated using super-time stepping \citep{Meyer_etal2012}, similarly to \citet{Johnston_etal2017}. The reduced mass is $\mu_m= m_f m_p$, where $m_p$ is the mass of proton and $m_f=1.2$. $k_B$ is the Boltzmann constant. 

The normalization is based on the photospheric values of density $\rho_\mathrm{u}=1.67 \times 10^{-7}\ \mathrm{g}\ \mathrm{cm}^{-3}$, length $H_\mathrm{u}=180 \ \mathrm{km}$ and magnetic field strength $B_\mathrm{u}=300 \ \mathrm{G}$. From these we obtain temperature $T_\mathrm{u}=6234~\mathrm{K}$, pressure $P_\mathrm{u}=7.16\times 10^3\ \mathrm{erg}\ \mathrm{cm}^{-3}$, velocity $v_\mathrm{u}=2.1\ \mathrm{km} \ \mathrm{s}^{-1}$ and time $t_\mathrm{u}=86.9\ \mathrm{s}$.

\begin{figure}
\begin{center}
\includegraphics[width=\columnwidth]{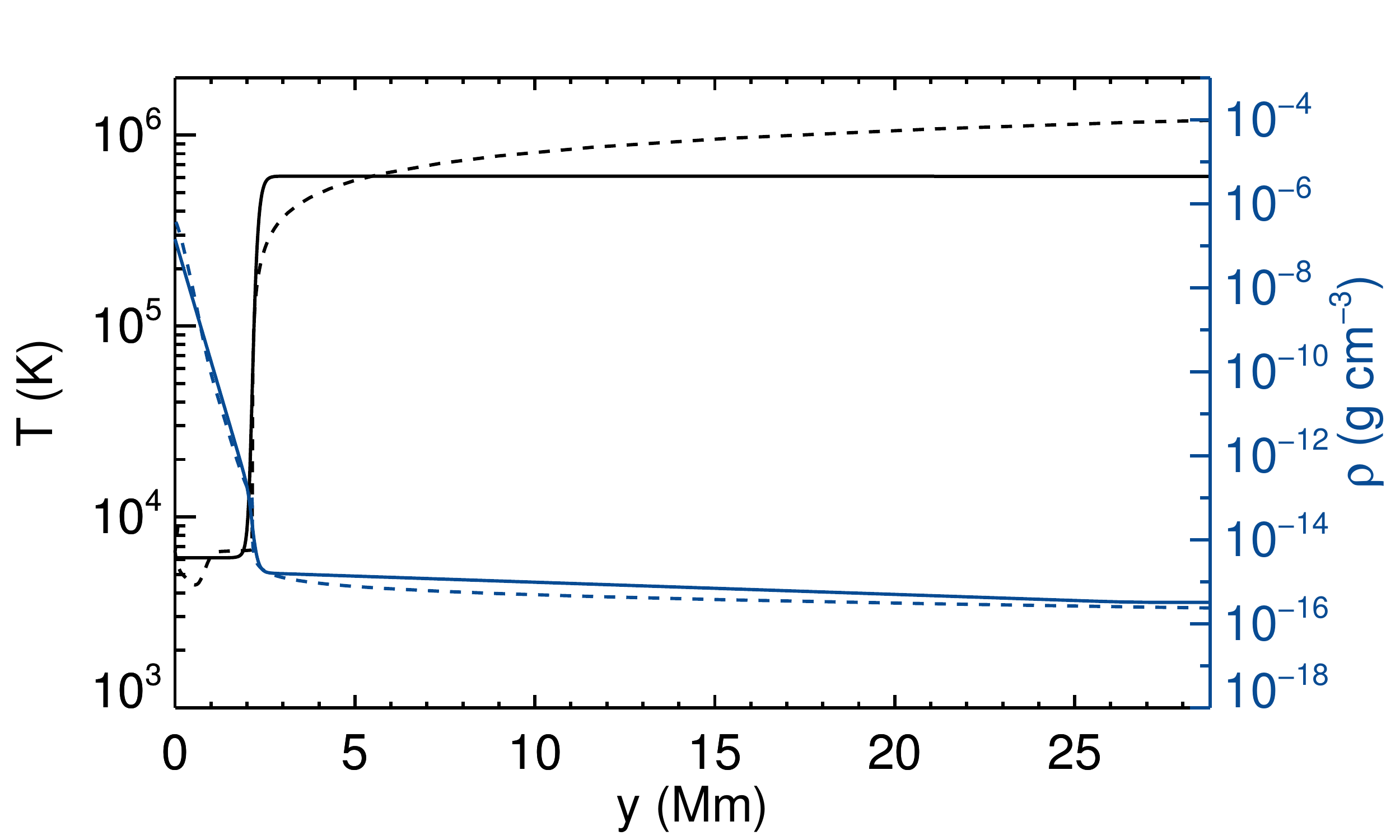}
\caption{
The atmospheric temperature (solid black) and density (solid blue). The dashed lines show the temperature and density of the 1D C7 model of \citet{Avrett_etal2008}.
}
\label{fig:stratification}
\end{center}
\end{figure}

The computational domain has a physical size of $x\in[-30,30]$~Mm in the horizontal direction and $y\in[0,30]$~Mm in the vertical direction, on a $2048\times1024$ uniform grid. The photosphere is at $y=0$.
To mimic the steep temperature increase from the photosphere to the corona, we assume an hyperbolic tangent profile for the atmospheric temperature
\begin{align}
    T(y) = T_{ph} + \frac{T_{cor}- T_{ph}}{2} 
            \left( \tanh{\frac{y-y_{cor}}{w_{tr}} +1} \right),
\end{align}
where $T_{ph}=6109$~K, $T_{cor} = 0.61$~MK, $y_{cor}=2.12$~Mm and $w_{tr}=0.18$~Mm. 
These parameters create an isothermal photospheric-chromospheric layer at $0 \ \mathrm{Mm} \le y < 1.96 \ \mathrm{Mm} $, a transition region  at $1.96 \ \mathrm{Mm} \le y < 3.3 \ \mathrm{Mm}$ and an isothermal coronal at $3.3 \ \mathrm{Mm} \le y < 30\ \mathrm{Mm}$.

To derive the atmospheric density, we assume the atmosphere is in hydrostatic equilibrium. We do so by numerically solving the hydrostatic equation $dP/dy = - gy$, assuming a photospheric density of $\rho_{ph}= 1.67\times10^{-7}$~g cm$^{-3}$.
The atmospheric temperature (solid black) and density  (solid blue) are shown in Fig.~\ref{fig:stratification}. For comparison, we plot with dashed lines the temperature and density for the 1D model atmosphere (model C7) of \citet{Avrett_etal2008}.

We adopt an anomalous resistivity
\begin{align}
    % \eta = 
      \eta=\begin{cases}
               \eta_0, \quad |j|<j_{crit} \\
               {\eta_0 + \eta_1}  |j|/j_{crit}. \quad |j|>j_{crit},
            \end{cases}
\end{align}
where $\eta_0=10^{-4}$, $\eta_1=10^{-3}$ and $j_{crit} = 10^{-3}$. The resistivity can be anomalous away from the boundaries ($x\in[-28,28]$~Mm and $y\in[2,28]$~Mm). Elsewhere, it is uniform with $\eta_0=10^{-4}$. Anomalous resistivity \citep{Yokoyama_etal1994} has been previously chosen to drive fast reconnection. Other methods \citep[e.g. hyper-diffusion][]{Nordlund_Stein_1990,vanBallegooijen_Cranmer_2008,Martinez_Sykora_etal2011} can also be used to initiate a fast reconnection by permitting enhanced resistivity in current sheets.

\begin{figure}
\begin{center}
\includegraphics[width=\columnwidth]{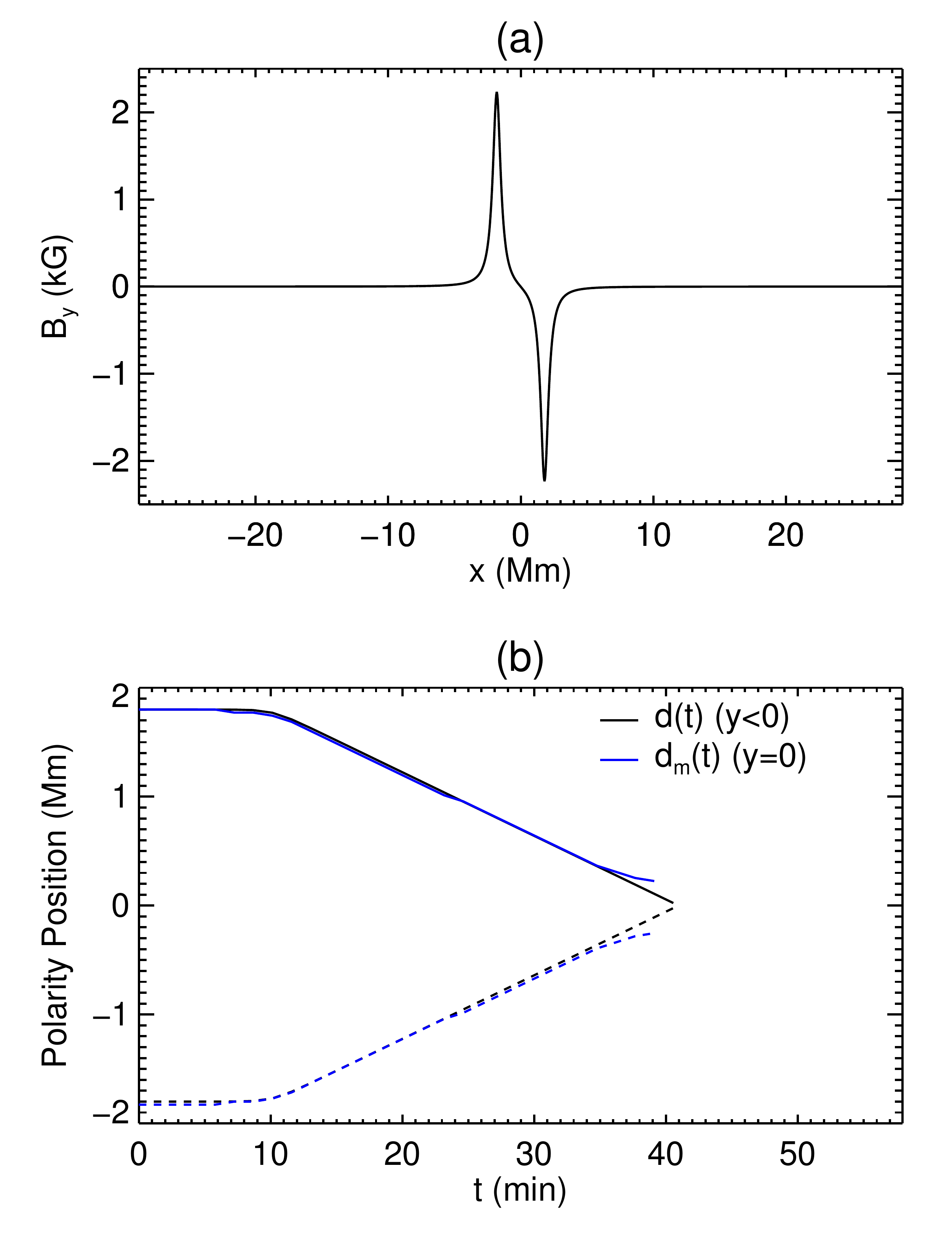}
\caption{
(a) The variation with $x$ of the vertical magnetic field ($B_y$) at the photosphere. (b) Black line: the position ($d$) of the sources as a function of time. Blue line: the position ($d_m$) of the photospheric polarities as a function of time.}
\label{fig:driver}
\end{center}
\end{figure}

The initial magnetic field is the sum of two magnetic sources and a horizontal field:
\begin{align}
    \mathbf{B} = \frac{F}{\pi}  \frac{ \mathbf{\hat{r}_1} }{r_1 }  - \frac{F}{\pi}  \frac{ \mathbf{\hat{r}_2} }{r_2} - B_0 \mathbf{\hat{x}},
    \label{eq:bfield_sim}
\end{align}
where
\begin{align}
    \mathbf{\hat{r}_1} = (x + d_s) \hat{x} + ( y - y_0) \hat{y}, \\
    \mathbf{\hat{r}_2} = (x - d_s) \hat{x} + ( y - y_0) \hat{y}
\end{align}
are the position vectors of the left and right sources, respectively, $d_s=1.8$~Mm is the distance of each source from $x=0$, and $y_0=-0.36$~Mm is the depth of the sources below the photosphere (the sources are outside the numerical domain).
The flux of each source is $F=2.5\times10^{11}$~Mx cm$^{-1}$. 
The polarities produced at the photosphere have a maximum field strength of 2.2~kG and a size of about 1 Mm (defined as the length where $|B_y|>100$~G) (Fig.~\ref{fig:driver}a).
The flux of each polarity is $F_{m} = 2.2\times10^{11}$~Mx cm$^{-1}$.
The horizontal field has a strength of $B_0= 45$~G. 

The boundary conditions on the upper boundary are $\mathbf{v}=0$ and zero gradients for $\mathbf{B}$, $\rho$, $\epsilon$. The photospheric boundary conditions are zero gradients for $\rho$, $\epsilon$. 
The magnetic field at the photospheric boundary changes according to the driver. 
To drive the cancellation, we move the sources with a velocity of 
\begin{align}
    v_0(t) = {\half v_{max}}  \left( \tanh{ \frac{t-t_0}{w} } +1  \right),
\end{align}
where $v_{max}$ = 1 \kms, $t_0=10.1$~min and $w=1.4$~min.
The positions of the sources changes according to $d(t) = d_s - x(t)$, where
\begin{eqnarray}
    x(t) = v_{max} \frac{w}{2} \left[  
                                      \ln{ \left( \cosh{ \frac{t-t_0}{w} } \right)} -
                                      \ln{ \left( \cosh{ \frac{t_0}{w} } \right)} 
                                \right]\\ \nonumber +
           \half v_{max} t. \ \ \ \ \ \ \ \ \ \ \ \ \ \ \ \ \ \ \ \ \ \ \ \ \ \ \ 
\end{eqnarray}
The simulation is driven by changing the magnetic field at the lower boundary ($y=0-$ (ghost cells)) using Equation~(\ref{eq:bfield_sim}) and $d(t)$.
The half-separation ($d(t)$) of the sources (below the photosphere) as a function of time  is plotted in Fig.~\ref{fig:driver}b (black line). The blue lines show the positions of the polarities at $y=0$ (found by measuring the location of maximum $B_y$). The latter reflects the response of the photosphere to the driver.

For the parametric study of Sec.~\ref{sec:parametric}, we vary the magnetic field strength ($B_0$) of the atmosphere  in order to vary the height of the null point. The values of $B_0$ and the corresponding null height at $t=0$~min are shown in Table~\ref{tab:table}.

\begin{deluxetable}{ccCrlc}[t]
\tablecaption{Initial Conditions for the Simulations \label{tab:table}}
\tablecolumns{3}
\tablenum{1}
\tablewidth{0pt}
\tablehead{
\colhead{Name} &
\colhead{$B_0$ (G)} &
\colhead{$y_N$ (Mm)}
}
\startdata
Case 1 &    45     & 7.6  & \\
Case 2 &    210    & 2.9  & \\
Case 3 &    300    & 2.2  & \\
Case 4 &    360    & 1.8  & \\
Case 5 &    600    & 0.9  & \\
\enddata
% \tablecomments{ 
% }
\end{deluxetable}

\subsection{Comparison of Theory with Simulation}
\label{sec:theory_comparison}

In this section, we discuss our reconnection experiment  driven by flux cancellation and compare its results with the theory presented in Sec.~\ref{sec:theory}. For this, we shall focus on Case 1 of Table~\ref{tab:table}.

\subsubsection{Brief Description of Simulation}

\begin{figure}
\begin{center}
\includegraphics[width=\columnwidth]{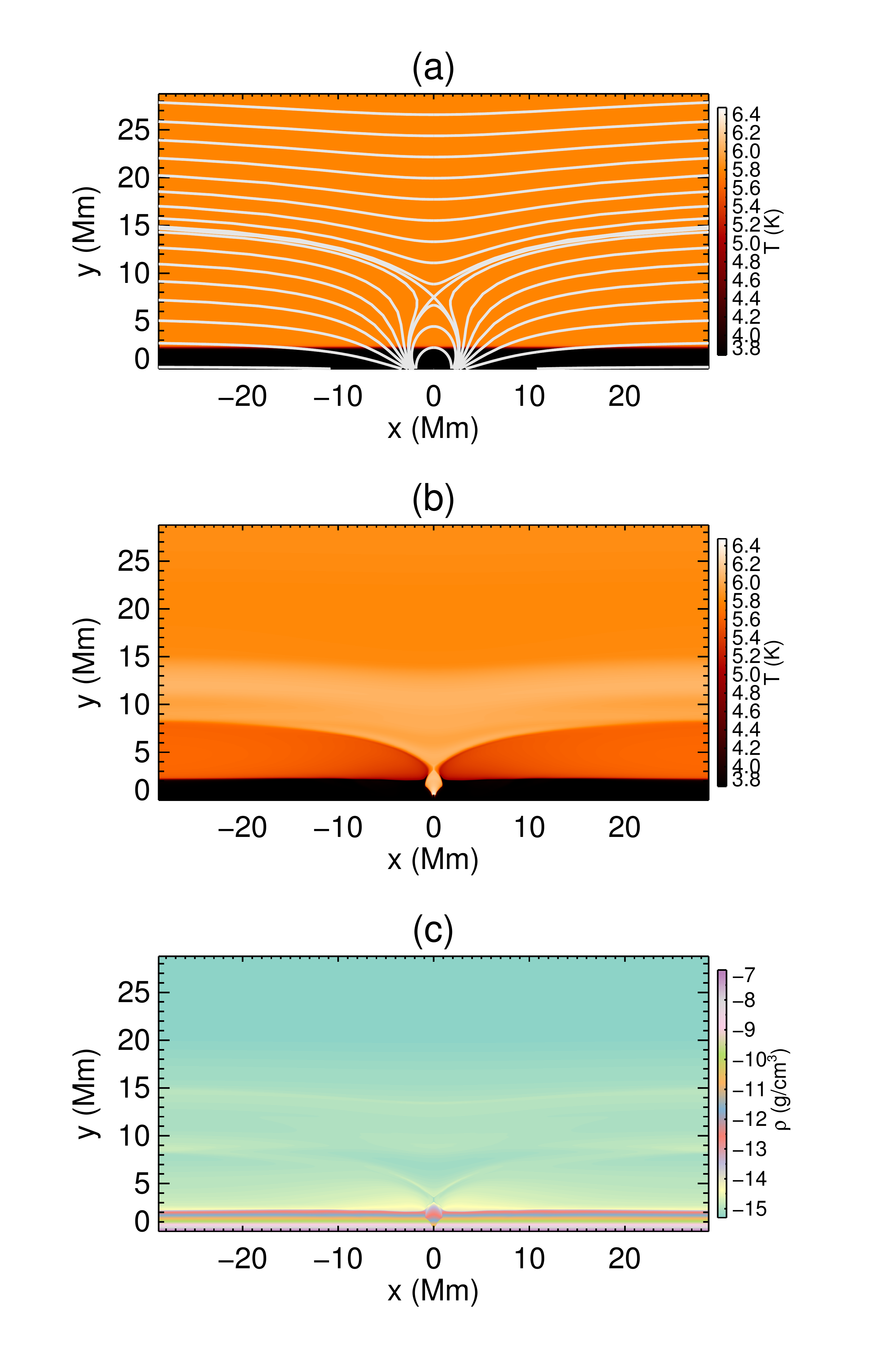}
\caption{ Case 1 simulation.
(a) Temperature and magnetic field lines at $t=0$.
(b) Temperature and (c) density at $t=40$~min.
}
\label{fig:temp_rho}
\end{center}
\end{figure}

The magnetic field at $t=0$ is shown in Fig.~\ref{fig:temp_rho}a, with a null point at $(x,y)=(0,7.6)$~Mm. As the driver is switched on, reconnection is driven at the null point due to the converging photospheric polarities. The energy released by reconnection spreads above and below the null (and shows up as a ``horizontal'' heated region and an underlying heated arcade in Fig.~\ref{fig:temp_rho}b). The heated material is denser than the background atmosphere (Fig.~\ref{fig:temp_rho}c).

At the photosphere, the positions of the polarities at $y=0$ ($d_m(t)$, blue line, Fig.~\ref{fig:driver}b) do not keep following the driver after $t=37$~min (black line). At this time, the magnitude of the photospheric field has decreased to the point that $\beta>1$. As a result, the driver cannot move the overlying field anymore.
The reconnection at the null follows the response of the atmospheric field to the driver and gradually stops. 
% Above the photosphere, after the driving stops (sources are located at $x=0$), there is still a remaining magnetic arcade, which becomes a O-loop.

The interaction distance for this simulation is $d_0=200$ based on the sources and $d_{0_m}= 2F_m/(\pi B_0) =173.2$ based on the photospheric polarities.

\subsubsection{Comparison Methodology}

In our simulation, we set the gradient of $\mathbf{B}$ to be zero at the boundaries (besides the photosphere). 
The reason for this is as follows. 
After flux cancellation, if the polarities are completely cancelled, the remaining atmospheric field ought to be a horizontal field with strength $B_0$. This cannot happen in a finite numerical domain, but only in a semi-infinite one.
% In order for that to happen, flux from the region above the null point has to be transferred outside the numerical domain.
% The zero gradient boundary conditions create this effect.
% They are ``straightening'' the fieldlines, and make the upper boundary behave like an open boundary for the flux. Therefore, flux is ``escaping'' from the upper boundary.
To achieve that in the simulation domain, we use a zero gradient boundary condition. This ``straightens'' the field lines, mimicing the effect we require.

In Sec.~\ref{sec:theory}, the inflow speed is found by taking into account the rate of change of flux and conservation of flux. 
To compare the simulation with  theory, we need to carefully calculate the fluxes inside the numerical domain.
The Appendix calculates how much flux should be found inside and outside the finite numerical domain, from which we deduce a flux correcting factor $f$ (Equation~\ref{eq:f}, Fig.~\ref{fig:correction}). This is used to multiply several quantities ($v_i \rightarrow f v_i$, $L_{sp}^2 \rightarrow f^2 L_{sp}$, $L^2 \rightarrow f L^2$, $dW/dt \rightarrow f^2 dW/dt$), since  the simulation uses a finite domain rather than a semi-infinite one (as on the Sun, for which $f\rightarrow1$).

In the simulation there is a difference between the driver (the sources below the photosphere) and the response of the photosphere to the driver (the polarities at the photosphere). 
Our theory uses observables (such as the separation of the polarities and the photospheric velocity) to predict the inflowing magnetic field and the energy release. 
To compare  theory with the computational experiment, we  use as ``observables'' two sets of values: 

(i) the values of the driver (such as $d$, $v$), and 

(ii) the values measured at the photosphere (or elsewhere), which mimic an actual observation. 

We will refer to the latter quantities with a subscript $m$. Thus, the half-separation of the sources is $d$, whereas the half-separation of the photospheric polarities is $d_m$.

\begin{figure*}
\begin{center}
\includegraphics[width=\textwidth]{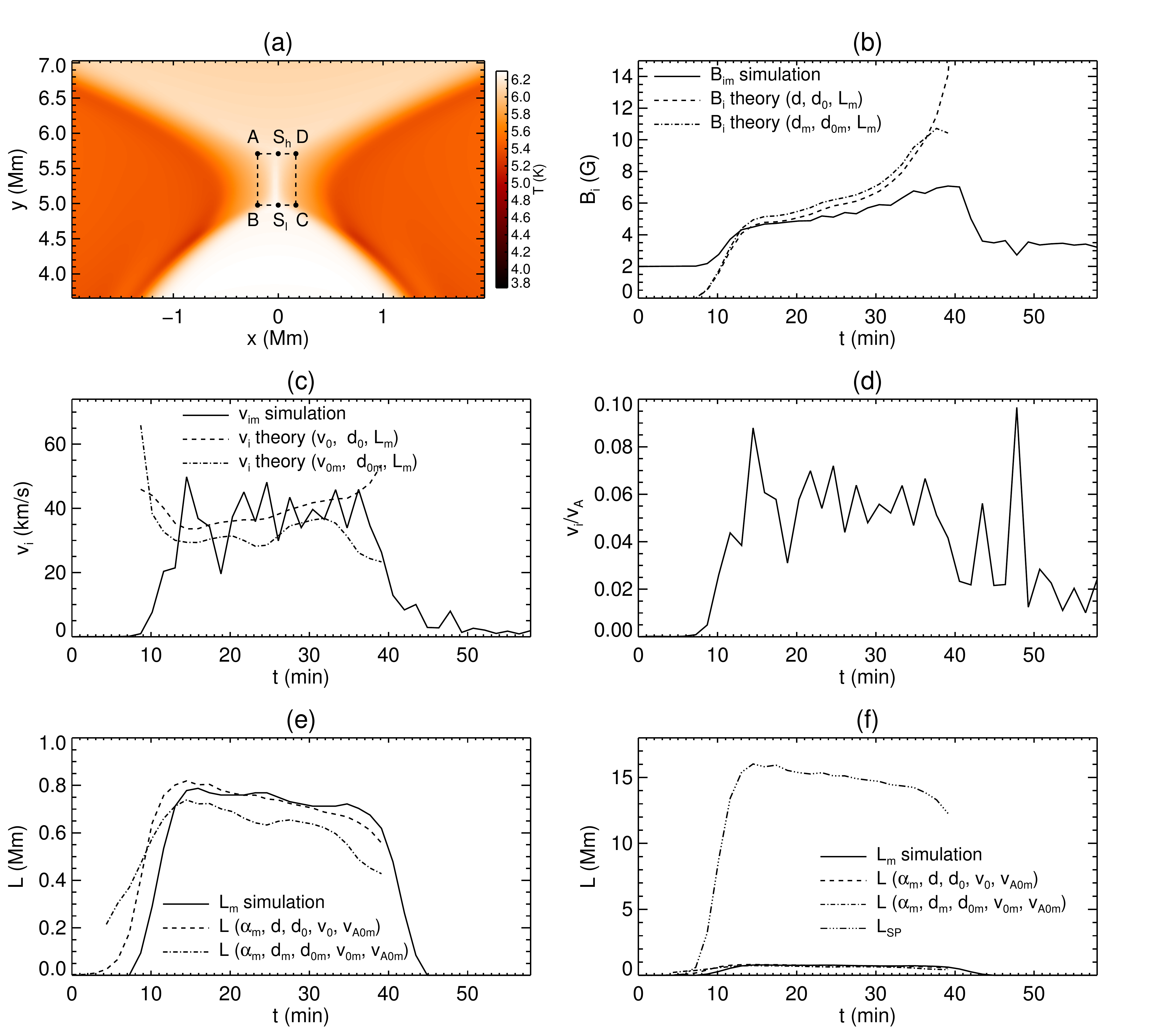}
\caption{
(a) Temperature around the reconnection site at $t=30$~min. 
Comparison between simulation and theory for 
(b) the inflow magnetic field,
(c) the inflow velocity,
(d) the Mach \alfven number of the inflow,
(e) the length of the current sheet,
(f) the length of the current sheet together with its value for Sweet-Parker reconnection.
}
\label{fig:inflow_var}
\end{center}
\end{figure*}

To compare the theory with the simulation, we first identify the current sheet. It is located along the $y$-axis  and is the vertical region of increased temperature located above the apex of the arcade (Fig.~\ref{fig:inflow_var}a, orange line segment). We identify the coordinates of its lowest ($S_{l}$) and highest ($S_{h}$) point throughout the simulation. The separation of these two points is the measured length of the current sheet, $L_m$ (solid black line, Fig.~\ref{fig:inflow_var}e).

We measure the inflowing magnetic field strength, velocity and density in the following manner. At both sides of the current sheet, we identify the regions that are parallel to the current sheet and at a distance of $\Delta x=0.2$~Mm away from it (line segments AB and CD, Fig.~\ref{fig:inflow_var}a). 
This distance is selected so that the current density there is at least an order of magnitude lower than the one inside the current sheet. 
We measure the values of these inflowing quantities as the average values of their magnitudes along both AB and CD, and so  find the average $B_{i_m}$, $v_{i_m}$ and $\rho_{i_m}$.

The total inflow of Poynting flux ($S_{i_m}$) into the current sheet is measured by taking into account the Poynting flux along both AC and CD. 

\subsubsection{Current Sheet Length, Inflowing Magnetic Field and Inflow Velocity}

Fig.~\ref{fig:inflow_var}b (solid line) shows the inflow magnetic field strength, $B_{i_m}$. The dashed line is the predicted $B_i$ using Equation~(\ref{eq:bi}) with $(d,d_0,L_m)$. The dashed-dot line is $B_i$ using Equation~(\ref{eq:bi}) with $(d_m,d_{0_m},L_m)$. Comparing both approaches, we see that the theory is in good agreement with the simulation. Indeed, the second approach, where we take into account only the response of the simulation to the driver, is in better agreement.

The simulation's inflow velocity ($v_{i_m}$) is plotted in Fig.~\ref{fig:inflow_var}c (solid line). The dashed line shows $v_i$ using Equation~(\ref{eq:vi}) (times $f$) with $(v_0,d_0,L_m)$. The dashed-dot line is $v_i$ using Equation~(\ref{eq:vi}) (times $f$) with $(v_{0_m},d_{0_m},L_m)$. Again, both agree well with the simulation.

Before estimating the length of the current sheet for fast reconnection (Equation~(\ref{eq:lfast})), we need to measure two quantities: $\alpha$, which is the  \alfven Mach number of the inflow and $v_{A0}$ (Equation~(\ref{eq:hybrid_va})). 
In Fig.~\ref{fig:inflow_var}d, we plot the  \alfven Mach number of the inflow. Between $t=10$ min and $40$~min, when the cancellation occurs, it has an average value of $\alpha_m=0.05$. This value of $\alpha$ is typical for fast  reconnection \citep{Priest_2014}.  
For the hybrid \alfven speed we use $v_{ {A0}_m } = B_0/\sqrt{\mu \rho_{i_m}}$.

The length of the simulation's current sheet ($L_m$) is plotted in Fig.~\ref{fig:inflow_var}e (solid line).
The dashed line is $L$ using Equation~(\ref{eq:lfast}) (times $f^2$) with $(\alpha_m,d,d_0,v_0, v_{ {A0}_m})$. 
The dashed-dot line is  $L$ using Equation~(\ref{eq:lfast}) (times $f^2$) with $(\alpha_m,d_m,d_{0_m},v_{0_m}, v_{ {A0}_m})$. Both approaches show the theory to be in good agreement with the simulation.

In Fig.~\ref{fig:inflow_var}f, we plot the quantities of panel e, and overplot the length of the current sheet assuming Sweet-Parker reconnection (triple-dot dashed line, with $L_{SP}$ calculated from Equation~(\ref{eq:lsp}) (times $f$) using $(d_{0_m},v_{0_m}, v_{ {A0}_m})$). 
The predicted current sheet for an assumption of slow Sweet-Parker reconnection is longer than the simulated one by an order of magnitude, and so we deduce that fast reconnection with an inflow \alfven Mach  speed of 0.05 well describes  the simulation.

\subsubsection{Energy Release}

\begin{figure}
\begin{center}
\includegraphics[width=\columnwidth]{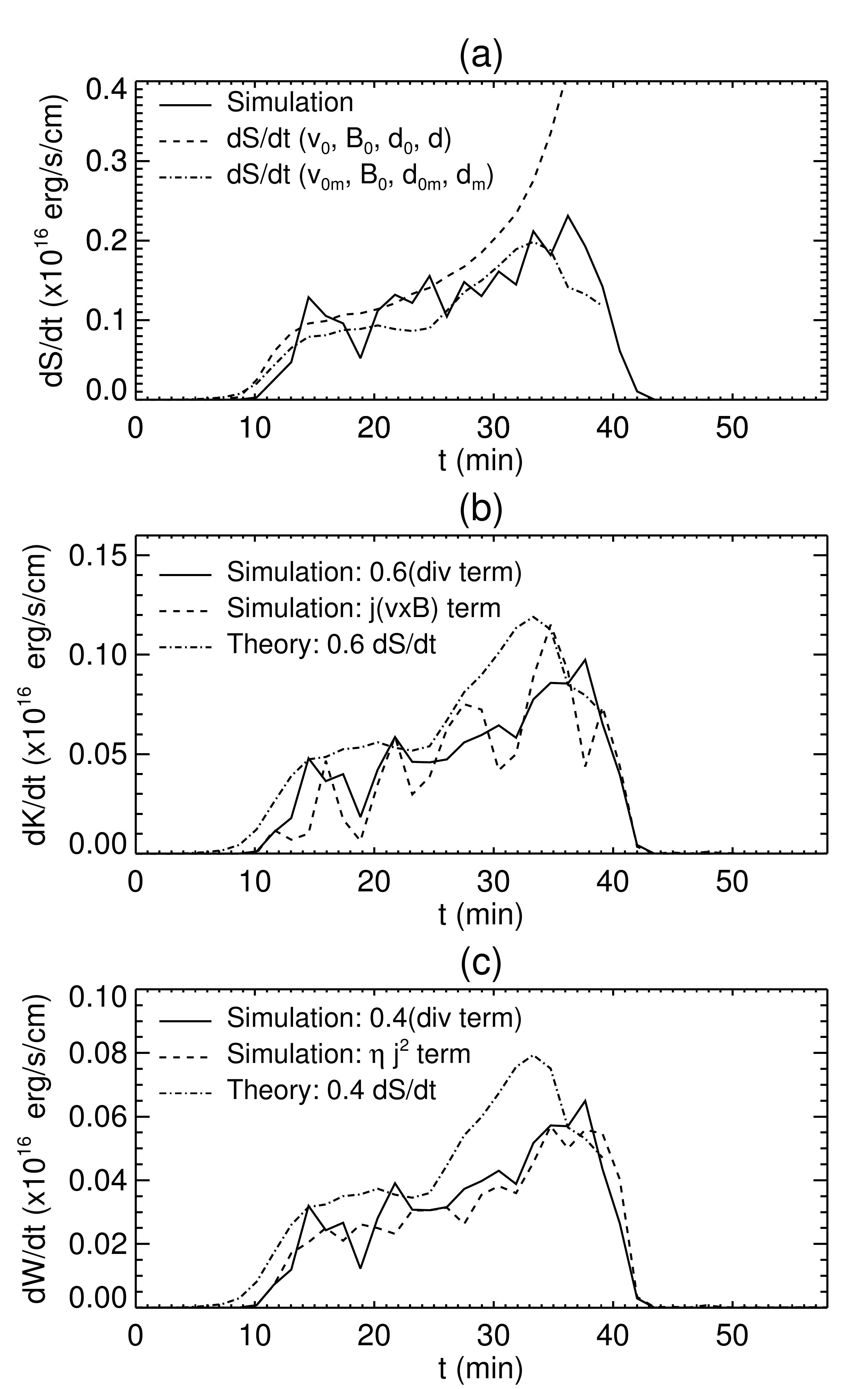}
\caption{ 
Comparison between simulation and theory for
(a) the total inflow of Poynting flux, 
(b) the rate of energy converted to kinetic energy and
(c) the rate of energy converted to heat during the reconnection.
}
\label{fig:energy}
\end{center}
\end{figure}

We study  energy release only for fast reconnection and first focus on the  Poynting flux inflow  ($S_{i_m}$), which is plotted in Fig.~\ref{fig:energy}a (solid lines). 
The dashed curve is $S_{i}$ from Equation~(\ref{eq:poyntingfast}) (times $f^2$) based on ($v_0$, $v_{{A0}_m}$, $B_0$, $d_0$, $d$). The dashed-dot curve is $S_{i}$ based on  ($v_{0_m}$, $v_{{A0}_m}$, $B_0$, $d_{0_m}$, $d_m$). Both agree well with the simulation. The approach of using only ``measured'' values is in excellent agreement with the simulation results.

Next, we consider the conversion of  Poynting flux to kinetic and thermal energy, for which we calculate the energy integral terms:
\begin{align}
    \int_C \frac{1}{\mu} {\bf E}\times {\bf B}\cdot d{\bf C}  
    &= -\int_A \eta \mathbf{j}^2 dA + \int_A \mathbf{j}\cdot \left( \mathbf{v} \times \mathbf{B} \right) dA \nonumber \\
    &- \int_A  \pder[]{t}\left( \frac{B^2}{2\mu}\right) dA.
\end{align}
The curve $C$ is ABCDA in Fig.~\ref{fig:inflow_var}a and the surface $A$ is its area.
The term $\int_C  {\bf E}\times {\bf B} \cdot d{\bf C} = \int_A {\bf \nabla} \cdot \mathbf{S}\ dA$, where $\mathbf{S}$ is the Poynting vector, is the rate of energy inflow and can be compared with Equation~(\ref{eq:poyntingfast}).
The term $-\int_A \eta \mathbf{j}^2 dA$ is the rate of energy converted to joule heating during  reconnection, which can be compared with Equation~(\ref{eq:heat}). 
The term $\int_A \mathbf{j}\cdot \left( \mathbf{v} \times \mathbf{B} \right) dA$ is the rate of energy converted to kinetic energy and can be compared with Equation~(\ref{eq:kinetic}). 
The term $-\int_A [1/(2\mu)] \p B^2/\p t\ dA$ is negligible.

We first check whether the energy conversion rates of the simulation agree with those of fast reconnection, i.e. whether $\frac{3}{5}$ of the total Poynting influx is converted to kinetic energy and $\frac{2}{5}$ is converted to Joule heating. 
If the conversion rates are such, then we should find in the simulation that
\begin{align}
\frac{3}{5}\int_A  \nabla \cdot \mathbf{S}~ dA= \int_A \mathbf{j}\cdot \left( \mathbf{v} \times \mathbf{B} \right) dA.
\end{align}
We plot these  terms in Fig.~\ref{fig:energy}b, from which it can be seen that the left (solid line) and right  (dashed line) terms are in agreement.
Furthermore, we examine if
\begin{align}
    \frac{2}{5} \int_A \nabla \cdot \mathbf{S}~dA = - \int_A \eta \mathbf{j}^2 dA.
\end{align}
These terms are plotted in Fig.~\ref{fig:energy}c, from which again the left  (solid line) and right  (dashed line) terms are in agreement.
So, indeed the energy release in the simulation agrees with the rates predicted by fast reconnection.

We now compare the energy release from the simulation with the theoretical predictions.
The kinetic energy release rate is calculated from  Equation~(\ref{eq:kinetic}) based on  ($v_{0_m}$, $v_{{A0}_m}$, $B_0$, $d_{0_m}$, $d_m$) and is  plotted in Fig.~\ref{fig:energy}b (dashed-dot line). This is in fact just the dashed-dot line of Fig.~\ref{fig:energy}a multiplied by 0.6. 
Next, we  calculate the total rate of conversion of energy to heat from  Equation~(\ref{eq:kinetic}) based on  ($v_{0_m}$, $v_{{A0}_m}$, $B_0$, $d_{0_m}$, $d_m$) and plot it Fig.~\ref{fig:energy}c (dashed-dot line). In both cases, the theoretical predictions are in excellent agreement with the simulation.

\begin{figure}
\begin{center}
\includegraphics[width=\columnwidth]{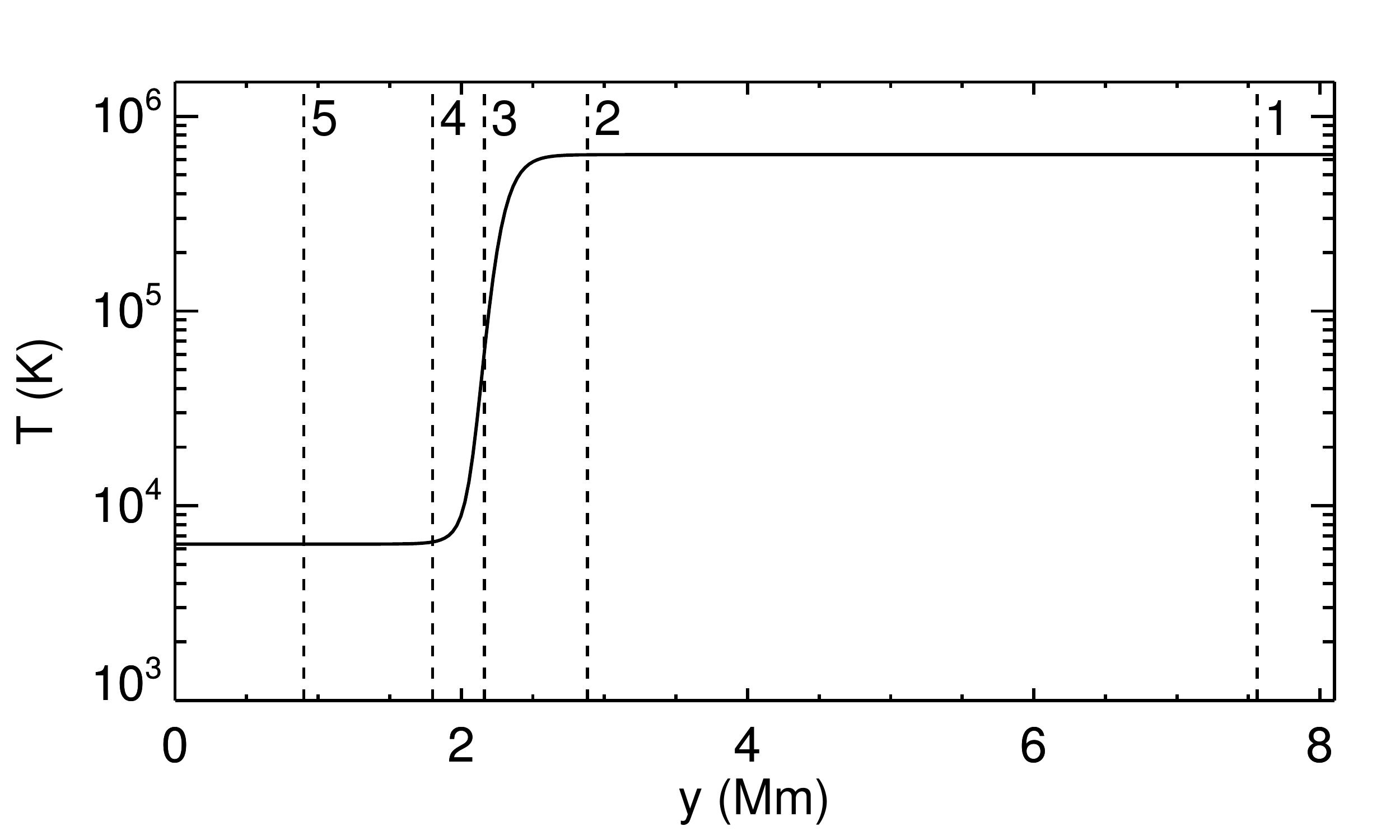}
\caption{ The vertical lines show the height of the null at $t=0$~min for all cases of Table~\ref{tab:table}, plotted against the background temperature stratification.}
\label{fig:null_height}
\end{center}
\end{figure}

\begin{figure*}
\begin{center}
\includegraphics[width=\textwidth]{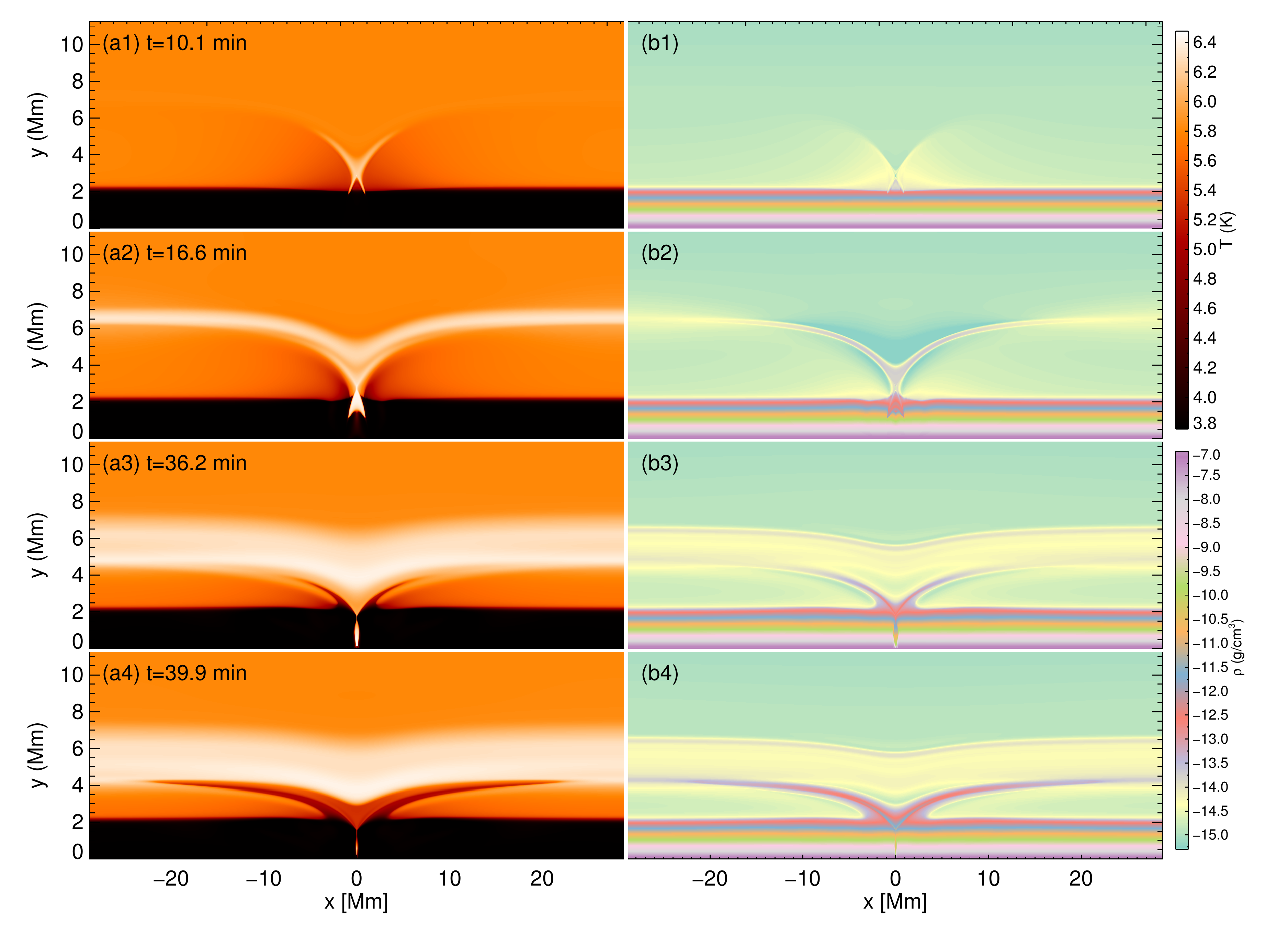}
\caption{ Evolution of temperature (left column) and density (right column) for case 2.}
\label{fig:case2}
\end{center}
\end{figure*}

\begin{figure*}
\begin{center}
\includegraphics[width=\textwidth]{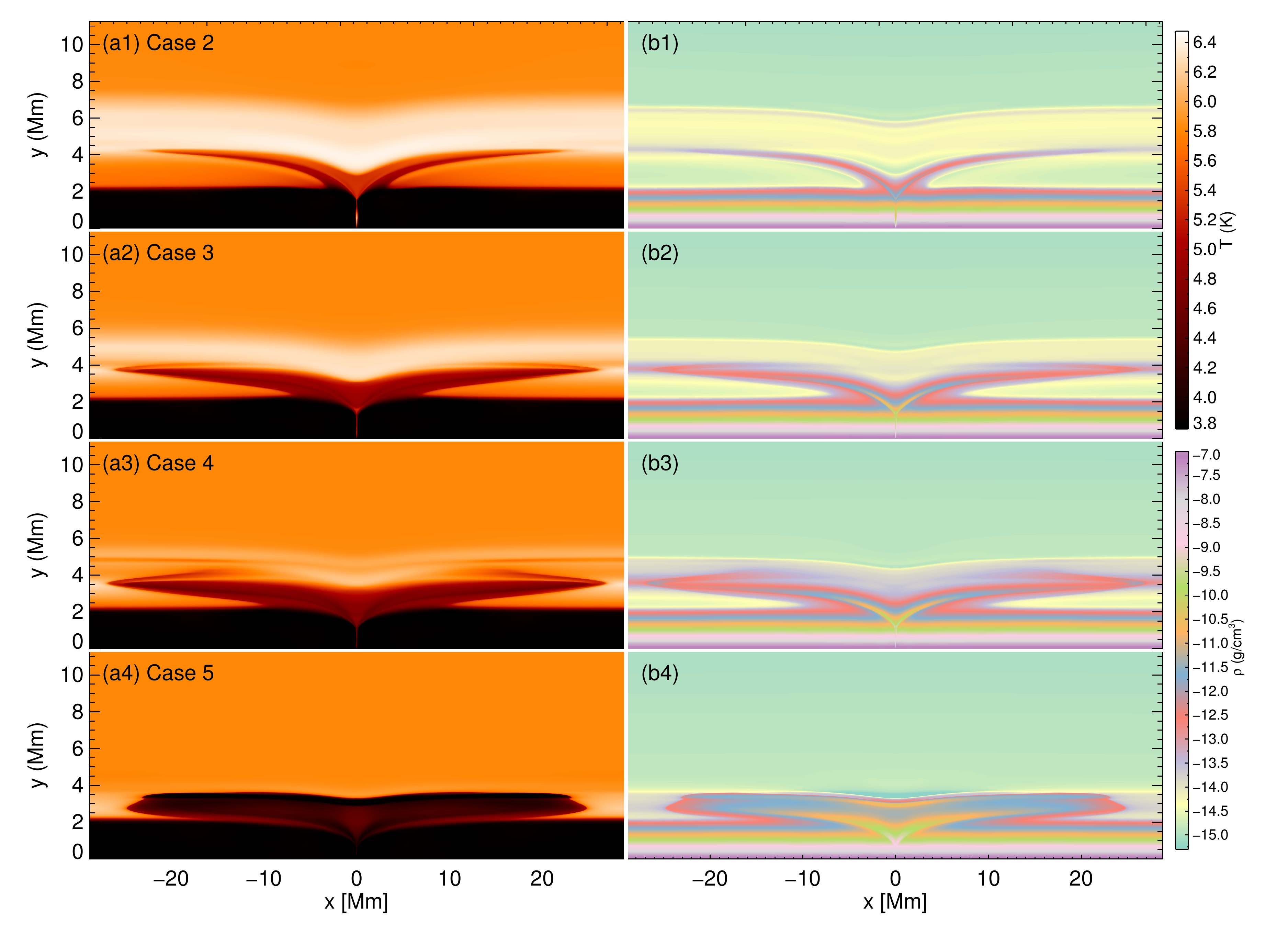}
\caption{ Temperature (left column) and density (right column) for cases 2-5 at $t=40$~min.}
\label{fig:parametric}
\end{center}
\end{figure*}

\subsection{Atmospheric Response}
\label{sec:parametric}

In this section we briefly discuss the   atmospheric response
to  reconnection driven by flux cancellation. 
First, we  study the time evolution of one individual case.
Then, we  vary the height of the null at $t=0$ by changing the strength of the external horizontal field and studying five cases with different $B_0$. The values of $B_0$ and the corresponding $y_N$ are shown in Table~\ref{tab:table}.
In Fig.~\ref{fig:null_height}, we plot the $y_N$ values (vertical lines) and the initial temperature stratification (solid line), in order to better visualize the initial location of the null point relative to the corona, transition region, chromosphere and photosphere.

We focus on the  time evolution of the temperature and density for case 2 (Fig.~\ref{fig:case2}). 
The null point is initially located at the base of the corona. 
As  reconnection starts, hot material is ejected along the post-reconnection field lines (panels a1, b1). A hot ``loop'' of 1.8~MK  and density of $2\times10^{-14}$~g~cm$^{-3}$ is formed above the reconnection site. Below the null, the top of the arcade is heated to 2.6~MK (panels a2, b2). 
As the polarities converge the null height  decreases, as predicted from the theory.
%(as $d<\half d_0$ in this case, see Fig.~\ref{fig:null_flux}a).
When the null point reaches the base of the transition region and below, dense, cool plasma is ejected along the reconnected field lines (Fig.~\ref{fig:case2}, panels a3, b3). 
Due to the higher density of the region, the resulting heat released from the reconnection cannot raise the plasma temperature to millions of Kelvin. As the process continues, a cooler and denser ejection is formed, with temperature of $0.05-0.12$~MK, and a density of $2\times10^{-12}-2\times10^{-13}$~g~cm$^{-3}$. It  propagates with velocity up to $105$~\kms, extending from transition region to coronal heights (panels a4, b4).

In Fig.~\ref{fig:parametric} we plot the temperature (first column) and density (second column) for cases $2-5$ at $t=40$~min, while Case 1 is shown in Fig.~\ref{fig:temp_rho}b,c. 
An important qualitative difference appears between the cases. 
When the null point is initially located in the corona, both a hot and a cool plasma region develop above the null during the cancellation. 
When the initial null point  is placed at progressively lower heights (top to bottom row), the amount of hot material decreases, while the cool material increases. 
Eventually, for a null point placed at the chromosphere (bottom row), the resulting post reconnection plasma does not have a high-temperature component.
In this case, the region above the null contains 

(i) a  very cool component of photospheric or chromospheric material (around 6300 K) which is ``slingshotted'' upwards from the tension of the reconnected lines with speed of 10-20~\kms and

(ii) a cool plasma component which is heated by reconnection to around $0.01-0.03$~MK. 

\begin{figure*}
\begin{center}
\includegraphics[width=\textwidth]{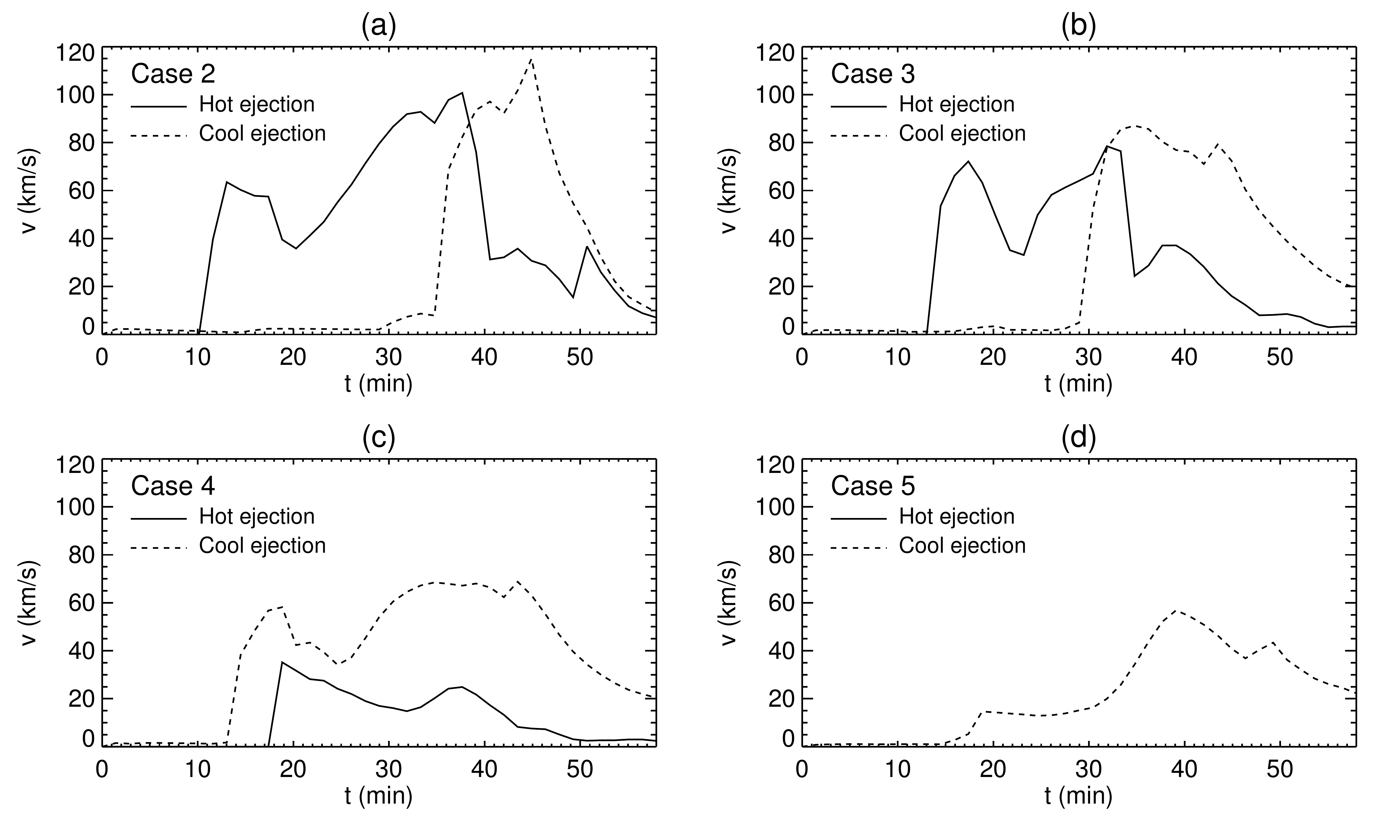}
\caption{ Maximum velocity of the hot ($T>1$~MK, solid lines) and cool ($T<0.2$~MK, dashed lines) plasma for (a) Case 2, (b) Case 3, (c) Case 4 and (d) Case 5.}
\label{fig:velocity}
\end{center}
\end{figure*}

In Fig.~\ref{fig:velocity} we plot the time evolution of the maximum velocity of the hot ($T>1$~MK) and cool ($T<0.2$~MK) plasma components for the cases shown in Fig.~\ref{fig:parametric}. The maximum velocities of the hot (cool) plasma ejections from (a) to (d) are 100~\kms (105~\kms), 79~\kms (87~\kms), 35~\kms (70~\kms) and 0~\kms (57~\kms)  respectively.
Notice also that the hot and cool components are produced with a time delay, as shown previously in Fig.~\ref{fig:case2}.
The time difference between the acceleration of the hot and cold material decreases as the null point is situated lower. In cases 2 and 3, the hot material appears first and the cold material after. For case 4, the hot and cold ejections are almost co-temporal.

\subsection{Atmospheric Response}
\label{sec:parametric}

\section{Discussion}

In \citet{Priest_etal2018}, we proposed magnetic reconnection driven by photospheric  flux cancellation as a mechanism for energizing coronal loops and heating the chromosphere. We also derived analytical expressions that predict the energy release during reconnection.
In the present work, we begin to numerically validate our theory by developing the theory in 2D and comparing it with computations of two converging polarities inside a stratified atmosphere containing a background horizontal field.
As the polarities converge, reconnection is driven at the null point. 

To compare the theory with simulations, we evaluated several quantities from the simulations. For example, we calculated the velocity of approach of the opposite polarities  in two ways. One was to use the values that correspond to the simulation's driver and the other was to measure the response of the photosphere to that driver.

We found  excellent agreement between  theory and  simulation, especially with the second approach. 
The response to the driver is to initiate motions in the numerical domain which lead to reconnection. 
It is found that the theory agrees well with the system's response to  the driver, which is encouraging, since it shows that
that our theory could be used to derive estimates of the energy released during flux cancellation from solar observations, as observations measure the photospheric and atmospheric response, without knowledge of the sub-photospheric conditions driving the cancellation.
We conclude, based on our 2D computational experiments, that the energy released during photospheric cancellation can be accurately estimated from a knowledge of  the converging velocity, the separation and strengths of the converging fluxes,  the strength of the background magnetic field, and the density and \alfven Mach  number of the material flowing into the current sheet.

The promising results from these 2D simulations suggest that our analytical estimates can indeed be used to predict  energy release. Ideal observational candidates for such a comparison in future include many cases where photospheric flux cancellation is associated with small-scale energy release, such as Ellerman bombs, UV bursts and IRIS bombs or the energy injected into coronal loops due to flux cancellation at their feet.

We have also presented an initial study of the atmospheric response to  reconnection (for more sophisticated and realistic simulations, see, e.g., \citet{Danilovic_etal2017,Hansteen_etal2017, nobrega18}). The maximum height of the null point in 2D is $\half d_0$. As the polarities converge, the null point moves up to its maximum height and then down towards the photosphere. The atmospheric response during photospheric cancellation is as follows.
When the null point is located initially at a coronal height, a hot ``loop'' (around $1-2$~MK) can be formed above the reconnection region. 
Cooler material is ejected along the reconnected field lines when the null point is located at the base of the transition region or below, with a velocity of 95~km s$^{-1}$ and a temperature of $0.05-0.1$~MK. These ejections occur with a time difference, which is smaller when the initial null point height is lower.
However, if the null point is initially located below the base of the transition region, no hot material is ejected, and we only find the formation of a cooler ejection. 
Thus, the location of energy release  is crucial for the type of plasma structure that is created. 
The hot structures that we find have temperatures and densities similar to those of a coronal loop, whereas the cool structures have values that are reminiscent of surges or larger spicules. 

Note that, if only part of the photospheric flux cancels, the null point  stops moving towards the photosphere at some intermediate height. 
Then, reconnection  occurs only between the initial and final height of the null, which  produces a shorter, less energetic burst of energy release.

There is an important difference between 2D and 3D simulations.
In 2D, the magnetic field of a source falls off with distance like $1/r$, whereas in 3D, the fall-off behaves like $1/r^2$. As a result, the interaction distance  in 2D ($d_0=2F/(\pi B_{0})$)  is larger than its 3D value ($d_0 = \sqrt{F/(\pi B_0)})$ \citep[see][]{Priest_etal2018}, which produces a higher location for 
 the null point in 2D  than in 3D for a given polarity separation distance, photospheric flux and background field.
Thus, in our 2D simulations, in order to place the null at a particular height in the stratified atmosphere, we adopt a stronger background field than would be needed in 3D. 
During the reconnection, this higher  background field produces a larger  Poynting influx into the current sheet in 2D than in 3D. 
The result is that more energy is converted into heat and kinetic energy. 
Therefore, we leave a detailed discussion of temperature and density distributions  along reconnected field lines for a future 3D experiment. 
In 3D, the energy release may well accelerate the cooler plasma to form shorter structures than in 2D.

In our model we have assumed a horizontal external field in order to be able to make a direct comparison with our analytical theory. 
An oblique external field would have several extra effects. 
Firstly, it would enhance plasma draining along field lines, changing the maximum length and density of the heated plasma structures, an effect that would be stronger for the cooler ejections. 
Secondly, the null point would  move sideways, as well as vertically. This could affect the width of the structures and possibly produce ``thread-like'' ejections. Thirdly, after reconnection, the flows above the null point will have up and down components, instead of being mainly horizontal. The resulting magnetic ``loop'' would have its footpoints rooted in the photosphere and the  ejection would be dominated by a single inclined upflow (together with a much shorter downflow), rather than consisting of two opposite directly horizontal flows  (e.g., Fig.~\ref{fig:case2}). Jet-like structures have been observed at the feet of coronal loops \citep{chitta17a,chitta17b} which could be related to the upflows we expect in the oblique field.
However, we do not expect the energy release to change drastically. In Sec.~\ref{sec:theory}, we derived the rate of heating by assuming it is half the total inflow of Poynting flux. For an oblique field, the flux function $\psi$ would be different, but, during flux cancellation, the same amount of flux will be cancelled, irrespective of the orientation of the external field. Consequently, the same  Poynting influx  into the current sheet would be produced over the same time scale. Thus, the energy release rate should not be significantly different. We aim to check this  numerically in future.

In this work, we have positively validated our analytical theory using 2D numerical computations. This suggests that nanoflares driven by magnetic flux cancellation can indeed be an important mechanism for heating the chromosphere and corona, as  proposed in \citet{Priest_etal2018}, which is built upon recent observational findings.
In future, we aim to extend our model in several ways to make it more realistic and to consider more cases. In particular, we shall study oblique external fields in order to determine in more detail the ways in which chromospheric and coronal loops may be heated by reconnection at their footpoints. We shall also set up a fully three-dimensional computation in order to  study the extent and implications of our theory and to deduce in more detail the atmospheric response to  energy release. 

\acknowledgements

L.P.C. received funding from the European Union's Horizon 2020 research and innovation programme under the Marie Sk\l{}odowska-Curie grant agreement No.\,707837. This research has made use of NASA's Astrophysics Data System.  The authors are most grateful for invaluable discussions with Hardi Peter, Clare Parnell and Alan Hood.

\appendix
\section{Flux Correction Factor}

At $x=0$, the flux contained between the heights $\bar{y}_1$ and $\bar{y}_2$ is $\psi = \int_{ \bar{y}_1}^{\bar{y}_2 } \bar{B}_x d\bar{y}$, which for our magnetic field (Equation~(\ref{eq:bfield_sim})) becomes:
\begin{align}
    \psi_{\bar{y}_1}^{\bar{y}_2} = \frac{2F}{\pi} 
                          \left[ \arctan{ \frac{\bar{y}_2-\bar{y}_0}{\bar{d}} } - (\bar{y}_2-\bar{y}_0) -
                                 \left(
                                     \arctan{ \frac{\bar{y}_1-\bar{y}_0}{\bar{d}} } - (\bar{y}_1-\bar{y}_0)
                                 \right)
                          \right].
\label{eq:flux_y1y2}                          
\end{align}
Thus, the total flux from the sources to the upper boundary ($y_{max}$) is
\begin{align}
    \psi_{\bar{y}_0}^{\bar{y}_{max}} = \frac{2F}{\pi} 
                         \arctan{ \frac{\bar{y}_{max}-\bar{y}_0}{\bar{d}} },
\label{eq:flux_y0ymax}
\end{align}
which depends on $\bar{d}$ and $\bar{y}_{max}$.
For a semi-infinite domain, such as considered in Sec.~\ref{sec:theory}, $\bar{y}_{max}\to\infty$ and so the flux becomes $F$, which is independent of $\bar{d}$. 
However, the simulation has a finite region, and so the dependence of  flux on $\bar{d}$ and $\bar{y}_{max}$ has to be taken into account in order to compare with  theory.

Consider  the fluxes above and below the null point.
The flux from the sources ($\bar{y}_0$) to the null ($\bar{y}^\prime_N= \sqrt{\bar{d} - \bar{d}^2} + \bar{y}_0)$ is:
\begin{align}
    \psi_{\bar{y}_0}^{\bar{y}_N^\prime} = \frac{2F}{\pi} 
                          \left( \arctan{ \frac{\bar{y}_N}{\bar{d}} } - \bar{y}_N 
                          \right),
\label{eq:flux_y0yn}                          
\end{align}
whereas the flux from the null point to the upper boundary of the numerical domain is
\begin{align}
    \psi_{\bar{y}_N^\prime}^{\bar{y}_{max}} = \frac{2F}{\pi} 
                          \left[ \arctan{ \frac{\bar{y}_{max}-\bar{y}_0}{\bar{d}} } - (\bar{y}_{max}-\bar{y}_0) -
                             \left(
                                 \arctan{ \frac{\bar{y}_N}{\bar{d}} } - \bar{y}_N 
                             \right)
                          \right].
\label{eq:flux_ynymax}
\end{align}
As the sources cancel (and $\bar{d}$ decreases from 1 to 0), the flux below the null point changes from 0 to $F$, resulting in a total cancelled flux of
\begin{align}
    \Delta\psi_{\bar{y}_0}^{\bar{y}_N^\prime} =F.
    \label{eq:flux_difference}
\end{align}
The flux above the null up to $\bar{y}_{max}$ changes by    
\begin{align}
    \Delta\psi_{\bar{y}_N^\prime}^{\bar{y}_{max}} = - \frac{2F}{\pi} \arctan{(\bar{y}_{max}-y_0)},
    \label{eq:excess_flux}
\end{align}
which becomes $\Delta\psi_{\bar{y}_N^\prime}^{\infty}=-F$ as $\bar{y}_{max}\to\infty$.
Therefore, in a semi-infinite domain, when moving the sources from $\bar{d}=1$ to $\bar{d}=0$,  there is flux balance between the fluxes below and above the null. 
However, for a finite $\bar{y}_{max}$,  there is  extra flux above $\bar{y}>\bar{y}_{max}$ which we do not take into account.
Thus, in a finite domain, it is not possible to fully cancel the two magnetic sources, to give a configuration with a uniform horizontal field, since  a flux of $|F|$ would be cancelled below the null while adding less than $-|F|$ above the null.

In the simulation, 
%we force the sources to fully cancel. Also, the boundary conditions (zero magnetic field gradient) are such that, if the field of the polarities fully cancels, the resulting atmospheric field will be a straight field of strength $B_0$ parallel to the photosphere, as expected from the theory. These boundary conditions ``straighten'' the field during cancellation, leading to the desired post-cancellation configuration. The resulting addition of flux to the system has to be taken into account when comparing  theory and simulation.
the rates of change of flux from the sources up to the null and from the null up to $\bar{y}_{max}$ are
\begin{align}
    \dot{\psi}_{\bar{y}_0}^{\bar{y}_N^\prime} =
                          \frac{2F}{\pi} \frac{v_0}{d_0} \sqrt{\frac{1}{\bar{d}}-1},
\label{eq:dpsi_y0yn}                          
\end{align}
and
\begin{align}
    \dot{\psi}_{\bar{y}_N^\prime}^{\bar{y}_{max}} = \dot{\psi}_f -\dot{\psi}_{\bar{y}_0}^{\bar{y}_N^\prime},
    \label{eq:dpsi_ynymax}
\end{align}
where 
\begin{align}
\dot{\psi}_f=\frac{2F}{\pi} \frac{v_0}{d_0}
                            \frac{ \bar{y}_{max}-\bar{y}_0 }{ (\bar{y}_{max}-\bar{y}_0)^2 + \bar{d}^2}.
\label{eq:dpsi_f}                            
\end{align}
The flux outside the finite domain changes at a rate $\dot{\psi}_{\bar{y}_{max}}^\infty =-\dot{\psi}_f$,
and the rate of change of flux added to the region above the null is $-|v_i b_i|$.
Therefore, below the null, from Equation~(\ref{eq:dpsi_ynymax}), the rate of change has to be:
\begin{align}
    \dot{\psi} \equiv \dot{\psi}_{\bar{y}_0}^{\bar{y}_N^\prime} =
        |v_i b_i| + \dot{\psi}_f
\label{eq:psi_dot_inflow_speed}
\end{align}
% in order for the simulations' flux to be balanced.
For a semi-infinite domain, $\dot{\psi}_f \to 0$, and therefore $\dot{\psi}_{\bar{y}_N^\prime}^{\bar{y}_{max}} =-\dot{\psi}_{\bar{y}_0}^{\bar{y}_N^\prime}$.

\begin{figure}
\begin{center}
\includegraphics[width=0.45\columnwidth]{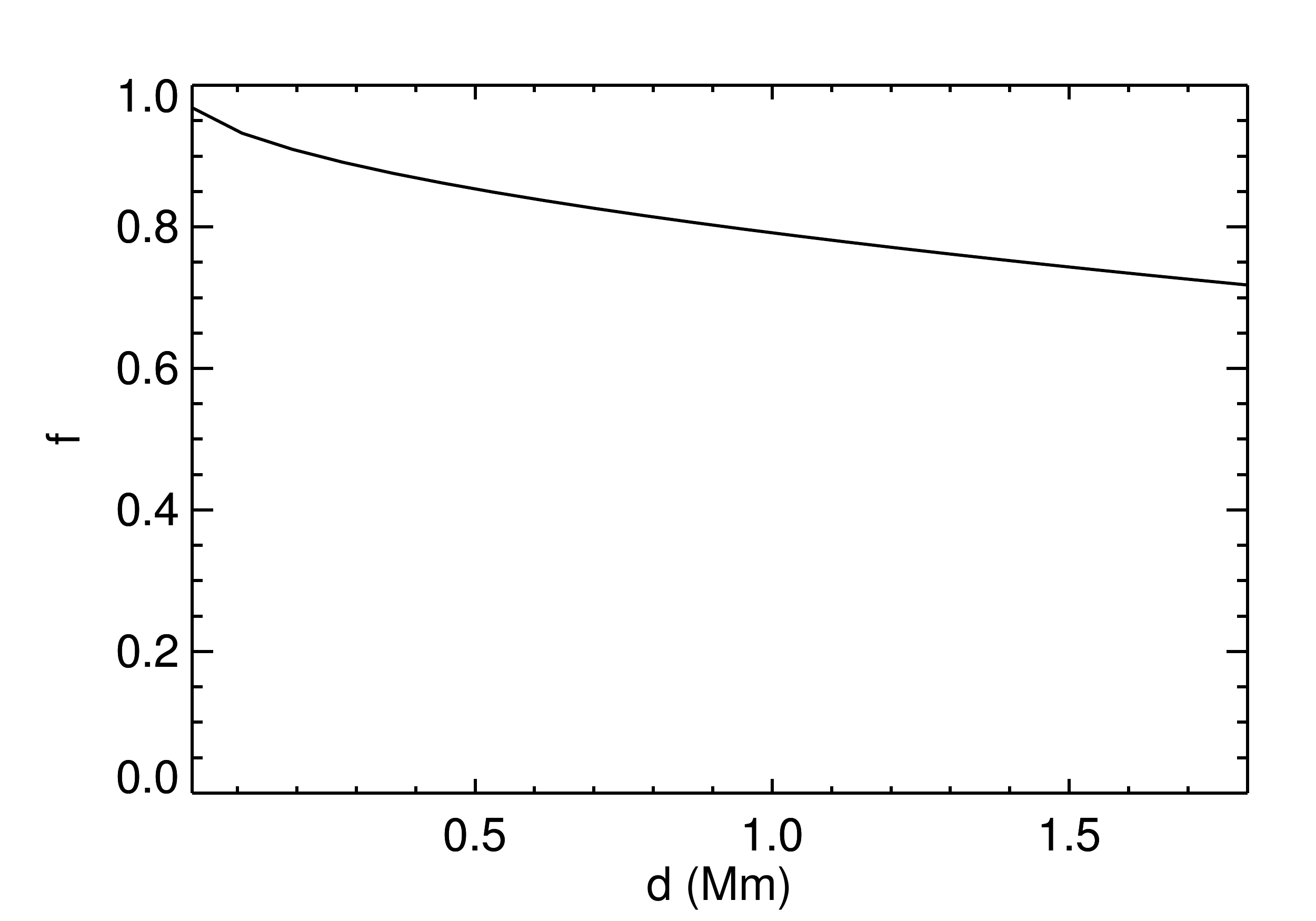}
\caption{
The flux correction factor ($f$) from Equation~(\ref{eq:f}), using the simulation's values for $\bar{d}(t)$ and $\bar{y}_{max}$.
}
\label{fig:correction}
\end{center}
\end{figure}

This affects the theory in the following way. 
In Sec.~\ref{sec:theory}, the inflow speed was found using the conservation of flux and $B_i$:
\begin{align}
v_i B_i = \dot{\psi}.
\end{align}
To compare  theory with simulation, we must use  $\dot{\psi}$ from Equation~(\ref{eq:psi_dot_inflow_speed}) to give
\begin{align}
    v_i = v_0 \frac{d_0}{L} 
            \left(
                1- \frac{ \bar{y}_{max}-\bar{y}_0 }{ (\bar{y}_{max}-\bar{y}_0)^2 + \bar{d}^2} 
                \frac{1}{ \sqrt{1/\bar{d} - 1} }
            \right)
\end{align}
or
\begin{align}
    v_i = f v_0 \frac{d_0}{L}, 
\label{eq:vi_correction}    
\end{align}
where 
\begin{align}
    f = 1- \frac{ \bar{y}_{max}-\bar{y}_0 }{ (\bar{y}_{max}-\bar{y}_0)^2 + \bar{d}^2} 
                \frac{1}{ \sqrt{1/\bar{d} - 1} }.
\label{eq:f}                
\end{align}
$f$ is a flux correction factor, which
 is plotted in Fig.~\ref{fig:correction}
 and which modifies several of the previous expressions, namely,
changing $L_{sp}^2 \rightarrow f^2 L_{sp}$, $L^2 \rightarrow f L^2$, ${dW}/{dt} \rightarrow f^2 {dW}/{dt}$. 
For a semi-infinite domain ($y_{max}\to \infty$), $f\rightarrow1$ and we recover the theory of Sec.~\ref{sec:theory} with $v_i \rightarrow v_0 {d_0}/{L}$.

\bibliographystyle{aasjournal} 
\bibliography{bibliography}

\begin{thebibliography}{}
\expandafter\ifx\csname natexlab\endcsname\relax\def\natexlab#1{#1}\fi
\providecommand{\url}[1]{\href{#1}{#1}}
\providecommand{\dodoi}[1]{doi:~\href{http://doi.org/#1}{\nolinkurl{#1}}}
\providecommand{\doeprint}[1]{\href{http://ascl.net/#1}{\nolinkurl{http://ascl.net/#1}}}
\providecommand{\doarXiv}[1]{\href{https://arxiv.org/abs/#1}{\nolinkurl{https://arxiv.org/abs/#1}}}

\bibitem[{{Arber} {et~al.}(2001){Arber}, {Longbottom}, {Gerrard}, \&
  {Milne}}]{Arber_etal2001}
{Arber}, T., {Longbottom}, A., {Gerrard}, C., \& {Milne}, A. 2001, {Journal of
  Computational Physics}, 171, 151 , \dodoi{10.1006/jcph.2001.6780}

\bibitem[{{Archontis} \& {Hood}(2010)}]{Archontis_etal2010}
{Archontis}, V., \& {Hood}, A.~W. 2010, \aap, 514, A56,
  \dodoi{10.1051/0004-6361/200913502}

\bibitem[{{Avrett} \& {Loeser}(2008)}]{Avrett_etal2008}
{Avrett}, E.~H., \& {Loeser}, R. 2008, \apjs, 175, 229, \dodoi{10.1086/523671}

\bibitem[{{Bhattacharjee} {et~al.}(2009){Bhattacharjee}, {Huang}, {Yang}, \&
  {Rogers}}]{bhattacharjee09}
{Bhattacharjee}, A., {Huang}, Y.~M., {Yang}, H., \& {Rogers}, B. 2009, Phys.
  Plasmas, 16, 112102

\bibitem[{Birn \& Priest(2007)}]{birn07}
Birn, J., \& Priest, E.~R. 2007, Reconnection of Magnetic Fields: MHD and
  Collisionless Theory and Observations (Cambridge, UK: Cambridge University
  Press)

\bibitem[{Birn {et~al.}(2001)Birn, Drake, Shay, Rogers, Denton, Hesse,
  Kuznetsova, Ma, Bhattacharjee, Otto, \& Pritchett}]{birn01}
Birn, J., Drake, J., Shay, M., {et~al.} 2001, J. Geophys. Res., 106, 3715,
  \dodoi{10.1029/1999JA900449}

\bibitem[{Biskamp(1986)}]{biskamp86}
Biskamp, D. 1986, Phys. Fluids, 29, 1520

\bibitem[{{Chitta} {et~al.}(2018){Chitta}, {Peter}, \& {Solanki}}]{chitta18}
{Chitta}, L.~P., {Peter}, H., \& {Solanki}, S.~K. 2018, \aap, 615, L9,
  \dodoi{10.1051/0004-6361/201833404}

\bibitem[{{Chitta} {et~al.}(2017{\natexlab{a}}){Chitta}, {Peter}, {Young}, \&
  {Huang}}]{chitta17a}
{Chitta}, L.~P., {Peter}, H., {Young}, P.~R., \& {Huang}, Y.-M.
  2017{\natexlab{a}}, Astron.\ Astrophys., 605, A49,
  \dodoi{10.1051/0004-6361/201730830}

\bibitem[{{Chitta} {et~al.}(2017{\natexlab{b}}){Chitta}, {Peter}, {Solanki},
  {Barthol}, {Gandorfer}, {Gizon}, {Hirzberger}, {Riethm{\"u}ller}, {van
  Noort}, {Blanco Rodr{\'{\i}}guez}, {Del Toro Iniesta}, {Orozco Su{\'a}rez},
  {Schmidt}, {Mart{\'{\i}}nez Pillet}, \& {Kn{\"o}lker}}]{chitta17b}
{Chitta}, L.~P., {Peter}, H., {Solanki}, S.~K., {et~al.} 2017{\natexlab{b}},
  Astrophys.\ J. Suppl., 229, 4, \dodoi{10.3847/1538-4365/229/1/4}

\bibitem[{{Danilovic} {et~al.}(2017){Danilovic}, {Solanki}, {Barthol},
  {Gandorfer}, {Gizon}, {Hirzberger}, {Riethm{\"u}ller}, {van Noort}, {Blanco
  Rodr{\'\i}guez}, {Del Toro Iniesta}, {Orozco Su{\'a}rez}, {Schmidt},
  {Mart{\'\i}nez Pillet}, \& {Kn{\"o}lker}}]{Danilovic_etal2017}
{Danilovic}, S., {Solanki}, S.~K., {Barthol}, P., {et~al.} 2017, The
  Astrophysical Journal Supplement Series, 229, 5,
  \dodoi{10.3847/1538-4365/229/1/5}

\bibitem[{Forbes \& Priest(1987)}]{forbes87}
Forbes, T., \& Priest, E. 1987, Rev. Geophys., 25, 1583,
  \dodoi{10.1029/RG025i008p01583}

\bibitem[{Golub {et~al.}(1974)Golub, Krieger, Silk, Timothy, \&
  Vaiana}]{golub74}
Golub, L., Krieger, A., Silk, J., Timothy, A., \& Vaiana, G. 1974, Astrophys.
  J., 189, L93

\bibitem[{{Hansteen} {et~al.}(2017){Hansteen}, {Archontis}, {Pereira},
  {Carlsson}, {Rouppe van der Voort}, \& {Leenaarts}}]{Hansteen_etal2017}
{Hansteen}, V.~H., {Archontis}, V., {Pereira}, T.~M.~D., {et~al.} 2017, \apj,
  839, 22, \dodoi{10.3847/1538-4357/aa6844}

\bibitem[{Harvey \& Martin(1973)}]{harvey73}
Harvey, K.~L., \& Martin, S.~F. 1973, Solar Phys., 32, 389

\bibitem[{Heyvaerts {et~al.}(1977)Heyvaerts, Priest, \& Rust}]{heyvaerts77}
Heyvaerts, J., Priest, E., \& Rust, D. 1977, Astrophys. J., 216, 123,
  \dodoi{10.1086/155453}

\bibitem[{{Hong} {et~al.}(2017){Hong}, {Ding}, \& {Cao}}]{Hong_etal2017}
{Hong}, J., {Ding}, M.~D., \& {Cao}, W. 2017, \apj, 838, 101,
  \dodoi{10.3847/1538-4357/aa671e}

\bibitem[{{Huang} {et~al.}(2018){Huang}, {Mou}, {Fu}, {Deng}, {Li}, \&
  {Xia}}]{huang18}
{Huang}, Z., {Mou}, C., {Fu}, H., {et~al.} 2018, Astrophys.\ J., 853, L26,
  \dodoi{10.3847/2041-8213/aaa88c}

\bibitem[{Huba(2003)}]{huba03}
Huba, J.~D. 2003, in Space Simulations, ed. M.~Scholer, C.~Dum, \& J.~B\"uchner
  (New York: Springer), 170--197

\bibitem[{{Huba} \& {Rudakov}(2004)}]{huba04}
{Huba}, J.~D., \& {Rudakov}, L.~I. 2004, Phys. Rev. Lett., 93, 175003,
  \dodoi{10.1103/PhysRevLett.93.175003}

\bibitem[{{Johnston} {et~al.}(2017){Johnston}, {Hood}, {Cargill}, \& {De
  Moortel}}]{Johnston_etal2017}
{Johnston}, C.~D., {Hood}, A.~W., {Cargill}, P.~J., \& {De Moortel}, I. 2017,
  \aap, 597, A81, \dodoi{10.1051/0004-6361/201629153}

\bibitem[{{Kim} {et~al.}(2015){Kim}, {Yurchyshyn}, {Bong}, {Cho}, {Cho}, {Lee},
  {Lim}, {Park}, {Yang}, {Ahn}, {Goode}, \& {Jang}}]{Kim_etal2015}
{Kim}, Y.-H., {Yurchyshyn}, V., {Bong}, S.-C., {et~al.} 2015, \apj, 810, 38,
  \dodoi{10.1088/0004-637X/810/1/38}

\bibitem[{Lee \& Fu(1986)}]{lee86a}
Lee, L.-C., \& Fu, Z. 1986, J. Geophys. Res., 91, 6807,
  \dodoi{10.1029/JA091iA06p06807}

\bibitem[{{Libbrecht} {et~al.}(2017){Libbrecht}, {Joshi}, {Rodr{\'\i}guez},
  {Leenaarts}, \& {Ramos}}]{Libbrecht_etal2017}
{Libbrecht}, T., {Joshi}, J., {Rodr{\'\i}guez}, J. d. l.~C., {Leenaarts}, J.,
  \& {Ramos}, A.~A. 2017, \aap, 598, A33, \dodoi{10.1051/0004-6361/201629266}

\bibitem[{Longcope(1998)}]{Longcope_1998}
Longcope, D.~W. 1998, Astrophys.\ J., 507, 433

\bibitem[{{Loureiro} {et~al.}(2012){Loureiro}, {Samtaney}, {Schekochihin}, \&
  {Uzdensky}}]{loureiro12}
{Loureiro}, N.~F., {Samtaney}, R., {Schekochihin}, A.~A., \& {Uzdensky}, D.~A.
  2012, Phys. Plasmas, 19, 042303, \dodoi{10.1063/1.3703318}

\bibitem[{{Loureiro} {et~al.}(2007){Loureiro}, {Schekochihin}, \&
  {Cowley}}]{loureiro07}
{Loureiro}, N.~F., {Schekochihin}, A.~A., \& {Cowley}, S.~C. 2007, Phys.
  Plasmas, 14, 100703

\bibitem[{{Loureiro} {et~al.}(2013){Loureiro}, {Schekochihin}, \&
  {Uzdensky}}]{loureiro13}
{Loureiro}, N.~F., {Schekochihin}, A.~A., \& {Uzdensky}, D.~A. 2013, Phys. Rev.
  E, 87, 013102, \dodoi{10.1103/PhysRevE.87.013102}

\bibitem[{Martin {et~al.}(1985)Martin, Livi, \& Wang}]{martin85}
Martin, S.~F., Livi, S., \& Wang, J. 1985, Astrophys. J., 38, 929

\bibitem[{{Mart{\'\i}nez-Sykora} {et~al.}(2011){Mart{\'\i}nez-Sykora},
  {Hansteen}, \& {Moreno- Insertis}}]{Martinez_Sykora_etal2011}
{Mart{\'\i}nez-Sykora}, J., {Hansteen}, V., \& {Moreno- Insertis}, F. 2011,
  \apj, 736, 9, \dodoi{10.1088/0004-637X/736/1/9}

\bibitem[{{Meyer} {et~al.}(2012){Meyer}, {Balsara}, \&
  {Aslam}}]{Meyer_etal2012}
{Meyer}, C.~D., {Balsara}, D.~S., \& {Aslam}, T.~D. 2012, \mnras, 422, 2102,
  \dodoi{10.1111/j.1365-2966.2012.20744.x}

\bibitem[{{Moore} {et~al.}(2010){Moore}, {Cirtain}, {Sterling}, \&
  {Falconer}}]{moore10}
{Moore}, R.~L., {Cirtain}, J.~W., {Sterling}, A.~C., \& {Falconer}, D.~A. 2010,
  Astrophys. J., 720, 757, \dodoi{10.1088/0004-637X/720/1/757}

\bibitem[{{Moreno-Insertis} \& {Galsgaard}(2013)}]{morenoinsertis13}
{Moreno-Insertis}, F., \& {Galsgaard}, K. 2013, Astrophys. J., 771, 20,
  \dodoi{10.1088/0004-637X/771/1/20}

\bibitem[{{Nelson} {et~al.}(2016){Nelson}, {Doyle}, \&
  {Erd{\'e}lyi}}]{Nelson_etal2016}
{Nelson}, C.~J., {Doyle}, J.~G., \& {Erd{\'e}lyi}, R. 2016, \mnras, 463, 2190,
  \dodoi{10.1093/mnras/stw2034}

\bibitem[{Nelson {et~al.}(2017)Nelson, Freij, Reid, Oliver, Mathioudakis, \&
  Erdélyi}]{Nelson_etal2017}
Nelson, C.~J., Freij, N., Reid, A., {et~al.} 2017, The Astrophysical Journal,
  845, 16

\bibitem[{{N{\'o}brega-Siverio} {et~al.}(2017){N{\'o}brega-Siverio},
  {Mart{\'{\i}}nez-Sykora}, {Moreno-Insertis}, \& {Rouppe van der
  Voort}}]{nobrega17}
{N{\'o}brega-Siverio}, D., {Mart{\'{\i}}nez-Sykora}, J., {Moreno-Insertis}, F.,
  \& {Rouppe van der Voort}, L. 2017, \apj, 850, 153,
  \dodoi{10.3847/1538-4357/aa956c}

\bibitem[{{N{\'o}brega-Siverio} {et~al.}(2018){N{\'o}brega-Siverio},
  {Moreno-Insertis}, \& {Mart{\'{\i}}nez-Sykora}}]{nobrega18}
{N{\'o}brega-Siverio}, D., {Moreno-Insertis}, F., \& {Mart{\'{\i}}nez-Sykora},
  J. 2018, \apj, 858, 8, \dodoi{10.3847/1538-4357/aab9b9}

\bibitem[{{Nordlund} \& {Stein}(1990)}]{Nordlund_Stein_1990}
{Nordlund}, {\AA}., \& {Stein}, R.~F. 1990, Computer Physics Communications,
  59, 119, \dodoi{10.1016/0010-4655(90)90161-S}

\bibitem[{Parnell \& Priest(1995)}]{parnell95}
Parnell, C.~E., \& Priest, E.~R. 1995, Geophys. Astrophys. Fluid Dyn., 80, 255,
  \dodoi{10.1080/03091929508228958}

\bibitem[{Peter {et~al.}(2014)Peter, Tian, Curdt, Schmit, Innes, De~Pontieu,
  Lemen, Title, Boerner, Hurlburt, Tarbell, Wuelser, Mart{\'\i}nez-Sykora,
  Kleint, Golub, McKillop, Reeves, Saar, Testa, Kankelborg, Jaeggli, Carlsson,
  \& Hansteen}]{Peter_etal2014}
Peter, H., Tian, H., Curdt, W., {et~al.} 2014, Science, 346,
  \dodoi{10.1126/science.1255726}

\bibitem[{Petschek(1964)}]{petschek64}
Petschek, H. 1964, in The Physics of Solar Flares (Washington: NASA Spec. Publ.
  SP-50), 425--439

\bibitem[{Priest(1986)}]{priest86b}
Priest, E. 1986, Mit. Astron. Ges., 65, 41

\bibitem[{{Priest}(2014)}]{Priest_2014}
{Priest}, E. 2014, {Magnetohydrodynamics of the Sun}

\bibitem[{Priest \& Forbes(1986)}]{priest86a}
Priest, E., \& Forbes, T. 1986, J. Geophys. Res., 91, 5579

\bibitem[{Priest {et~al.}(1994)Priest, Parnell, \& Martin}]{priest94b}
Priest, E., Parnell, C., \& Martin, S. 1994, Astrophys. J., 427, 459

\bibitem[{{Priest} {et~al.}(2018){Priest}, {Chitta}, \&
  {Syntelis}}]{Priest_etal2018}
{Priest}, E.~R., {Chitta}, L.~P., \& {Syntelis}, P. 2018, \apjl, 862, L24,
  \dodoi{10.3847/2041-8213/aad4fc}

\bibitem[{{Reid} {et~al.}(2016){Reid}, {Mathioudakis}, {Doyle}, {Scullion},
  {Nelson}, {Henriques}, \& {Ray}}]{Reid_etal2016}
{Reid}, A., {Mathioudakis}, M., {Doyle}, J.~G., {et~al.} 2016, \apj, 823, 110,
  \dodoi{10.3847/0004-637X/823/2/110}

\bibitem[{{Rezaei} \& {Beck}(2015)}]{Rezaei_2015}
{Rezaei}, R., \& {Beck}, C. 2015, \aap, 582, A104,
  \dodoi{10.1051/0004-6361/201526124}

\bibitem[{{Rutten}(2016)}]{Rutten_2016}
{Rutten}, R.~J. 2016, \aap, 590, A124, \dodoi{10.1051/0004-6361/201526489}

\bibitem[{Rutten {et~al.}(2015)Rutten, van~der Voort, \&
  Vissers}]{Rutten_etal2015}
Rutten, R.~J., van~der Voort, L. H. M.~R., \& Vissers, G. J.~M. 2015, The
  Astrophysical Journal, 808, 133

\bibitem[{Shay \& Drake(1998)}]{shay98a}
Shay, M.~A., \& Drake, J.~F. 1998, Geophys. Res. Lett., 25, 3759,
  \dodoi{10.1029/1998GL900036}

\bibitem[{{Shay} {et~al.}(2007){Shay}, {Drake}, \& {Swisdak}}]{shay07}
{Shay}, M.~A., {Drake}, J.~F., \& {Swisdak}, M. 2007, Phys. Rev. Lett., 99,
  155002

\bibitem[{Shibata {et~al.}(1992)Shibata, Ishido, Acton, Strong, Hirayama,
  Uchida, McAllister, Matsumoto, Tsuneta, Shimizu, Hara, Sakurai, Ichimoto,
  Nishino, \& Ogawara}]{shibata92}
Shibata, K., Ishido, Y., Acton, L.~W., {et~al.} 1992, Publ. Astron. Soc. Japan,
  44, L173

\bibitem[{{Shimojo} \& {Shibata}(2000)}]{shimojo00}
{Shimojo}, M., \& {Shibata}, K. 2000, Astrophys. J., 542, 1100,
  \dodoi{10.1086/317024}

\bibitem[{{Smitha} {et~al.}(2017){Smitha}, {Anusha}, {Solanki}, \&
  {Riethm{\"u}ller}}]{smitha17}
{Smitha}, H.~N., {Anusha}, L.~S., {Solanki}, S.~K., \& {Riethm{\"u}ller}, T.~L.
  2017, Astrophys. J. Suppl., 229, 17, \dodoi{10.3847/1538-4365/229/1/17}

\bibitem[{{Solanki} {et~al.}(2010){Solanki}, {Barthol}, {Danilovic}, {Feller},
  {Gandorfer}, {Hirzberger}, {Riethm{\"u}ller}, {Sch{\"u}ssler}, {Bonet},
  {Mart{\'{\i}}nez Pillet}, {del Toro Iniesta}, {Domingo}, {Palacios},
  {Kn{\"o}lker}, {Bello Gonz{\'a}lez}, {Berkefeld}, {Franz}, {Schmidt}, \&
  {Title}}]{solanki10a}
{Solanki}, S.~K., {Barthol}, P., {Danilovic}, S., {et~al.} 2010, Astrophys. J.
  Letts., 723, L127

\bibitem[{{Solanki} {et~al.}(2017){Solanki}, {Riethm{\"u}ller}, {Barthol},
  {Danilovic}, {Deutsch}, {Doerr}, {Feller}, {Gandorfer}, {Germerott}, {Gizon},
  {Grauf}, {Heerlein}, {Hirzberger}, {Kolleck}, {Lagg}, {Meller}, {Tomasch},
  {van Noort}, {Blanco Rodr{\'{\i}}guez}, {Gasent Blesa}, {Balaguer
  Jim{\'e}nez}, {Del Toro Iniesta}, {L{\'o}pez Jim{\'e}nez}, {Orozco Suarez},
  {Berkefeld}, {Halbgewachs}, {Schmidt}, {{\'A}lvarez-Herrero},
  {Sabau-Graziati}, {P{\'e}rez Grande}, {Mart{\'{\i}}nez Pillet}, {Card},
  {Centeno}, {Kn{\"o}lker}, \& {Lecinski}}]{solanki17b}
{Solanki}, S.~K., {Riethm{\"u}ller}, T.~L., {Barthol}, P., {et~al.} 2017,
  Astrophys. J. Supplement, 229, 2, \dodoi{10.3847/1538-4365/229/1/2}

\bibitem[{{Syntelis} {et~al.}(2015){Syntelis}, {Archontis}, {Gontikakis}, \&
  {Tsinganos}}]{Syntelis_etal2015}
{Syntelis}, P., {Archontis}, V., {Gontikakis}, C., \& {Tsinganos}, K. 2015,
  \aap, 584, A10, \dodoi{10.1051/0004-6361/201423781}

\bibitem[{{Tian} {et~al.}(2016){Tian}, {Xu}, {He}, \& {Madsen}}]{Tian_etal2016}
{Tian}, H., {Xu}, Z., {He}, J., \& {Madsen}, C. 2016, \apj, 824, 96,
  \dodoi{10.3847/0004-637X/824/2/96}

\bibitem[{{Tiwari} {et~al.}(2014){Tiwari}, {Alexander}, {Winebarger}, \&
  {Moore}}]{tiwari14}
{Tiwari}, S.~K., {Alexander}, C.~E., {Winebarger}, A.~R., \& {Moore}, R.~L.
  2014, Astrophys.\ J. Letts., 795, L24, \dodoi{10.1088/2041-8205/795/1/L24}

\bibitem[{{Toriumi} {et~al.}(2017){Toriumi}, {Katsukawa}, \&
  {Cheung}}]{Toriumi_etal2017}
{Toriumi}, S., {Katsukawa}, Y., \& {Cheung}, M. C.~M. 2017, \apj, 836, 63,
  \dodoi{10.3847/1538-4357/836/1/63}

\bibitem[{{van Ballegooijen} \& {Cranmer}(2008)}]{vanBallegooijen_Cranmer_2008}
{van Ballegooijen}, A.~A., \& {Cranmer}, S.~R. 2008, \apj, 682, 644,
  \dodoi{10.1086/587457}

\bibitem[{van~der Voort {et~al.}(2017)van~der Voort, Pontieu, Scharmer, de~la
  Cruz~Rodríguez, Martínez-Sykora, Nóbrega-Siverio, Guo, Jafarzadeh,
  Pereira, Hansteen, Carlsson, \& Vissers}]{RouppevanderVoort2017}
van~der Voort, L.~R., Pontieu, B.~D., Scharmer, G.~B., {et~al.} 2017, The
  Astrophysical Journal Letters, 851, L6

\bibitem[{{Vissers} {et~al.}(2015){Vissers}, {Rouppe van der Voort}, {Rutten},
  {Carlsson}, \& {De Pontieu}}]{Vissers_etal2015}
{Vissers}, G.~J.~M., {Rouppe van der Voort}, L.~H.~M., {Rutten}, R.~J.,
  {Carlsson}, M., \& {De Pontieu}, B. 2015, \apj, 812, 11,
  \dodoi{10.1088/0004-637X/812/1/11}

\bibitem[{Vissers {et~al.}(2013)Vissers, van~der Voort, \&
  Rutten}]{Vissers_etal2013}
Vissers, G. J.~M., van~der Voort, L. H. M.~R., \& Rutten, R.~J. 2013, The
  Astrophysical Journal, 774, 32

\bibitem[{{Watanabe} {et~al.}(2011){Watanabe}, {Vissers}, {Kitai}, {Rouppe van
  der Voort}, \& {Rutten}}]{Watanabe_etal2011}
{Watanabe}, H., {Vissers}, G., {Kitai}, R., {Rouppe van der Voort}, L., \&
  {Rutten}, R.~J. 2011, \apj, 736, 71, \dodoi{10.1088/0004-637X/736/1/71}

\bibitem[{{Yokoyama} \& {Shibata}(1994)}]{Yokoyama_etal1994}
{Yokoyama}, T., \& {Shibata}, K. 1994, \apjl, 436, L197, \dodoi{10.1086/187666}

\bibitem[{{Yokoyama} \& {Shibata}(1996)}]{yokoyama96}
---. 1996, Pub. Astron. Soc Japan, 48, 353

\end{thebibliography}

%\end{article}
\end{document}